\documentclass[fleqn,usenatbib]{mnras}

\usepackage{newtxtext,newtxmath}

\usepackage[T1]{fontenc}
\usepackage{ae,aecompl}

\usepackage{graphicx}	
\usepackage{amsmath}	

\usepackage{amssymb}	

\def\la{\mathrel{\mathpalette\fun <}}
\def\ga{\mathrel{\mathpalette\fun >}}
\def\fun#1#2{\lower3.6pt\vbox{\baselineskip0pt\lineskip.9pt
        \ialign{$\mathsurround=0pt#1\hfill##\hfil$\crcr#2\crcr\sim\crcr}}}

\title[Roman galaxy redshift survey] {Linear bias and halo occupation distribution of emission line galaxies from Nancy Grace Roman Space Telescope}

\author[Z. Zhai et al.]{
Zhongxu Zhai,$^{1}$\thanks{E-mail: zhai@ipac.caltech.edu}
Yun Wang,$^{1}$
Andrew Benson,$^{2}$
Chia-Hsun Chuang,$^{3}$
Gustavo Yepes$^{4,5}$
\\
$^{1}$IPAC, California Institute of Technology, Mail Code 314-6, 1200 E. California Blvd., Pasadena, CA 91125 \\
$^{2}$Carnegie Observatories, 813 Santa Barbara Street, Pasadena, CA 91101 \\
$^{3}$Kavli Institute for Particle Astrophysics and Cosmology, Stanford University, 452 Lomita Mall, Stanford, CA 94305 \\
$^{4}$Departamento de F\'isica Te\'{o}rica, M\'{o}dulo 8, Facultad de Ciencias, Universidad Aut\'{o}noma de Madrid, 28049 Madrid, Spain \\
$^{5}$CIAFF, Facultad de Ciencias, Universidad Aut\'{o}noma de Madrid, 28049 Madrid, Spain \\
}
\date{Accepted XXX. Received YYY; in original form ZZZ}

\pubyear{2015}

\hypersetup{draft}
\begin{document}
\label{firstpage}
\pagerange{\pageref{firstpage}--\pageref{lastpage}}
\maketitle

\begin{abstract}

We present measurements of the linear galaxy bias of H$\alpha$ and [OIII] emission line galaxies (ELGs) for the High Latitude Spectroscopic Survey (HLSS) of Nancy Grace Roman Space Telescope, using galaxy mocks constructed using semi-analytical model for galaxy formation, {\it Galacticus}, with a large cosmic volume and redshift coverage. We compute the two-point statistics of galaxies in configuration space and measure linear bias within scales of $10\sim50 h^{-1}$Mpc. We adopt different selection algorithms to investigate the impact of the Roman line flux cut, as well as the effect of dust model used to calibrate Galacticus, on the bias measurements. We consider galaxies with H$\alpha$ and [OIII] emissions over the redshift range $1<z<3$, as specified by the current baseline for the Roman HLSS. We find that the linear bias for the H$\alpha$ and [OIII] ELGs can be expressed as a linear function with respect to redshift:
$b \simeq 0.88z+0.49$ for H$\alpha$ $(1<z<2)$, and  
$b \simeq 0.98z+0.49$ for [OIII] $(2<z<3)$. 
We have also measured the Halo Occupation Distributions of these H$\alpha$ and [OIII] emission line galaxies to understand their distribution within dark matter halos.
Our results provide key input to enable the reliable forecast of dark energy and cosmology constraints from Roman.

\end{abstract}


\section{Introduction}

As biased tracers of the underlying dark matter distribution, galaxies form primarily in the peaks of the matter density field. Therefore they are not uniformly distributed in the universe. At large scales, the distribution of galaxies reveals a coherent structure in the background of a cosmic web and this large scale structure depends on the fundamental cosmological parameters and the physics governing the formation and evolution of galaxies. The connection of the distribution between galaxies and dark matter can be described by galaxy bias, $b$, which can be obtained by comparing the clustering amplitudes of galaxies in the mock galaxy catalog and the dark matter simulation for a given cosmological model (\citealt{Coil_2013}). 
The proper modeling of galaxy bias is critical in facilitating the use of galaxy clustering as a cosmological probe.

Galaxy clustering data have been used to advance our understanding of both cosmology and galaxy formation. Retrieval of the information on large scales has been extensively studied with the linear perturbation theory of cosmic density field. The use of galaxy catalogs from spectroscopic redshift surveys has enabled the observations of large scale structure, which provides measurements for cosmic distance scales through the baryon acoustic oscillation (BAO), and the linear growth rate through the redshift space distortion (RSD) effect over a wide redshift range, see e.g. \cite{Eisenstein_2005, Cole_2005, Beutler_2011, Beutler_2012, Ross_2015, Delubac_2015, Ata_2018, Bautista_2018} and references therein. These measurements have been used to put constraints on the fundamental cosmological parameters. However, due to the lack of statistical precision and systematic accuracy, alternative theories to explain the cosmic acceleration, a.k.a. dark energy, are not conclusively ruled out. The minimal extension to the standard model, the $\Lambda$CDM cosmology, is allowed by the current observational data. In order to distinguish competing theories and constrain the parameter space, future galaxy surveys are required to probe cosmic large scale structure over wider redshift ranges and larger cosmic volumes  (\citealt{Wang_2008a, Wang_2008b}). 

For future galaxy surveys like Euclid (\citealt{Laureijs_2011, Laureijs_2012}), and NASA's Nancy Grace Roman Space Telescope (hereafter Roman, \citealt{Green_2012, Dressler_2012, Spergel_2015}), galaxy clustering will be one of the main cosmological probes used to measure the properties of dark energy and constrain possible deviations of gravity from general relativity. Roman will mainly target H$\alpha$ and [OIII] emission line galaxies (ELGs) within redshift $1.0<z<3.0$, complementary to Euclid by design. To maximize the science return of space missions, it is necessary to optimize survey strategies. In \cite{Merson_2018, Zhai_2019MNRAS}, we calibrated and applied a semi-analytical model (SAM) of galaxy formation, Galacticus (\citealt{Benson_2012}), to N-body simulation to produce a realistic synthetic galaxy catalog and estimate the number densities of H$\alpha$ and [OIII] emitters. Using the same SAM, we have produced a H$\alpha$ galaxy mock catalog for the Roman HLSS (\citealt{Zhai_2020}), to facilitate the development of analysis tools for Roman BAO/RSD science. We measured the clustering signal and adopted a theoretical template for galaxy power spectrum to investigate the significance of the BAO and RSD measurements. This simulated catalog also enables an estimate of the galaxy bias. 
This is crucial for future surveys like Roman in evaluating whether we can infer the properties of dark matter correctly, and forecast the power of the survey to constrain dark energy. \cite{Merson_2019} combined the SAM and halo occupation distribution (HOD) approach to produce a H$\alpha$ galaxy catalog and predict the linear bias as a function of redshift for both Roman and Euclid. In this work, we carry out a more precise analysis by using the simulated galaxy catalog from SAM only, to forecast the linear bias of both H$\alpha$ and [OIII] ELGs and their HODs, over the entire redshift range of $1 < z < 3$ for the Roman HLSS. Our results can also be used for additional tests for the underlying SAM. 

The bias relation between the distribution of galaxies and underlying matter has been extensively studied in literature. Euclid and Roman will be the first cosmological surveys targeting H$\alpha$ and [OIII] emission line galaxies. The detailed analysis of the H$\alpha$ and [OIII] emission line galaxies using either numerical or semi-analytical method can inform both cosmological and galaxy evolution studies. \cite{Nusser_2020} adopts various SAMs and empirical model for galaxy formation to investigate the biasing relation for an Euclid-like survey. The bias measurement at linear scale reveals a constant function of star formation rate (SFR) for star forming galaxies. By utilizing the luminosity variation and peculiar velocity field from the galaxy distribution in redshift space, the ELGs could provide a measurement of the linear growth rate without being biased by the environmental effects (\citealt{Nusser_2020}).

Since Euclid and Roman target ELGs at $z \ga 1$, their sample selection differs significantly from ground projects targeting lower redshift galaxies. It is important to investigate the connection of these ELGs and their host dark matter halos (\citealt{Wechsler_2018}). 
Using the observational data from the eBOSS ELG program and mock catalogs, \cite{Avila_2020} studied a series of models for Halo Occupation Distribution (HOD) of the ELGs and investigated the impact on the clustering measurement. Using similar observational data, \cite{Guo_2019} measured the occupancy of the star formation galaxies and the evolution as a function of stellar mass within redshift range $0.7<z<1.2$. Although the eBOSS ELGs have different redshift distribution and selection algorithm than Roman and Euclid, their implications for the galaxy properties can provide reference information for the future surveys. For other investigations of the connection between ELGs and dark matter halos, see, e.g. \cite{Hadzhiyska_2020, Jimenez_2020} and references therein. The connection of the Roman ELGs with the host dark matter halos is not only useful for cosmology and galaxy science, the modeling of their HODs can provide a convenient way to populate galaxies within a dark matter simulation while preserving the clustering properties. This enables the production of many mock galaxy catalogs in a fast and practical manner, required for constructing the covariance matrix for the likelihood analysis (\citealt{Norberg_2009}), and has been widely used in the literature (\citealt{CMASS_Martin, Zehavi_2011, Manera_2013, Zhai_2017}). 
The detailed investigation for the Roman galaxy redshift survey, presented in this work, is able to provide more details of how the dark matter halos are populated by galaxies with different star formation history and emission lines. 

The SAM calibrated and used in our work has enabled the estimate of the number density of ELGs as a function of redshift (\citealt{Zhai_2019MNRAS}). Along with the linear bias measurement in the current analysis, they provide the crucial input information to forecast the wide range of possible dark energy and cosmological science from Roman galaxy redshift survey, e.g., using the Fisher matrix approach (\citealt{Tegmark_1997}). 
This can serve as a convenient method to predict the constraining power on the properties of dark energy, for instance a Figure-of-merit analysis (\citealt{Wang_2008b}). In addition,
our results are useful in investigating the extra constraining power from galaxy bispectrum (\citealt{Yankelevich_2019}), the constraint on neutrino masses (\citealt{Hamann_2012}) and so on.

Our paper is organized as follows: in Section 2, we introduce the galaxy mock catalog for the Roman galaxy redshift survey and the selection algorithm of the sample. Section 3 presents the bias measurements from galaxy clustering, and the HOD of galaxies within the Roman redshift range. Finally we discuss and conclude in Section 4.

\section{Methodology}

In this section, we describe the construction of the simulated Roman catalogs of ELGs, and the sample selection for the Roman galaxy redshift survey.

\subsection{Galaxy formation model}

The synthetic galaxy catalog used in this paper has been constructed using the Galacticus galaxy formation model (\citealt{Benson_2012}). Similar to the other SAMs, Galacticus parametrizes the astrophysical processes and performs the evolution of galaxy populations within a distribution of dark matter halos and their merger trees. The processes governing galaxy formation and evolution include gas cooling, star formation, feedback from supernovae, black hole formation and so on. By parametrizing these processes as ordinary differential equations (ODE) and calling the ODE solver, Galacticus can perform a simulation of galaxies within a sufficiently large volume in a timely manner and output details for the galaxy populations, including the star formation history, galaxy morphology, spectral energy distribution (SED), photometric luminosities for a set of filter transmission curves and emission line luminosities. 

Before using Galacticus to produce the galaxy catalog, we need to determine the free parameters in the model due to the poor prior knowledge of the astrophysical processes. This can be non-trivial since the typical number of free parameters is 15 or more (\citealt{Wechsler_2018}). In our work, we do not limit ourselves to local galaxies to calibrate the model, but compare the model prediction with galaxy populations at higher redshifts relevant to Roman. The parameters of this model have been calibrated in \cite{Zhai_2019MNRAS}, including the parameters for the physics of galaxy formation and the dust-attenuation model. In particular, the dust model is calibrated to produce consistent prediction of H$\alpha$ luminosity function compared with observations from the ground-based narrow-band High-z Emission Line Survey (HiZELS, \citealt{Geach_2008, Sobral_2009, Sobral_2013}), or the number counts data collected from Wide Field Camera 3 (WFC3) Infrared Spectroscopic Parallels survey (WISP; \citealt{Atek_2010, Atek_2011, Mehta_2015}). The dust model applied is the \citet{Calzetti_2000} model with parameter $A_{V}$ to describe the strength of dust attenuation. The result is $A_{V}=1.652$ for calibration based on WISP, and $A_{V}=1.916$ for HiZELS. Note that higher value of $A_{V}$ means stronger dust attenuation, thus the HiZELS-based calibration results a lower number density of the galaxy sample compared with the WISP-based calibration, with a preference of selecting brighter galaxies. In this paper, we present the bias measurement for ELGs with both dust models and investigate the impact on the large scale structure analysis.

The galaxy population in this analysis is selected by the emission line luminosity. Galacticus can output the number of ionizing photons for various species (HI, He I and [OII]), the metallicity of the interstellar medium (ISM), the hydrogen gas density and the volume filling factor of HII regions. We use these parameters to interpolate the tabulated libraries from the CLOUDY photo-ionization code (\citealt{Ferland_2013}) and compute the emission line luminosity for each galaxy. More details of the method can be found in \citealt{Merson_2018}.

\subsection{N-body simulation}

The key ingredient for SAMs like Galacticus is the set of merger trees of dark matter halos, which can be approximately constructed using the Press-Schechter formalism (\citealt{PS_1974}), or come from a cosmological N-body simulation. Here we have chosen the later for high fidelity, and use the merger trees extracted from the UNIT simulation \footnote{\url{https://unitsims.ft.uam.es}} (\citealt{Chuang_2019}) which assumes a spatially flat $\Lambda$CDM model with parameters consistent with Planck 2016 measurement (\citealt{Planck_2016}). The simulation contains 4096$^3$ particles with a box-size of $1h^{-1}$Gpc. This simulation has a mass resolution of $10^{9}h^{-1}M_{\odot}$ with data product covering redshift range of $0<z<99$. The large volume and high resolution makes this simulation sufficient for the next generation galaxy surveys, including DESI, and those planned for Roman and Euclid. We refer the readers to \cite{Chuang_2019} for more details of the UNIT simulation and its data products. The merger trees of the dark matter halos are constructed using the Consistent Trees software (\citealt{Behroozi_2013b}) and the final product contains more than 160 million merger trees. Applying Galacticus to this simulation allows us to build a light cone catalog of galaxies. In particular, we use the method from \cite{Kitzbichler_2007} to determine where the dark matter halos enter the lightcone of the observer. The resulting catalog has an area of $\sim2000\text{deg}^{2}$, consistent with the current baseline design of Roman HLSS. 

\subsection{Luminosity function of H$\alpha$ and [OIII] emission lines}

\begin{figure*}
\begin{center}
\includegraphics[width=18.5cm, height=7cm]{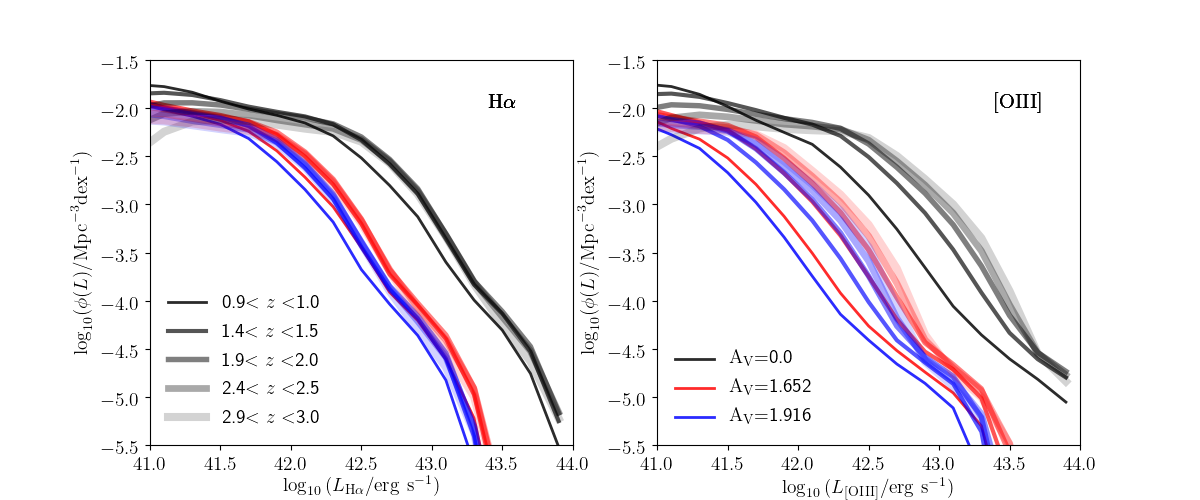}
\includegraphics[width=18.5cm, height=7cm]{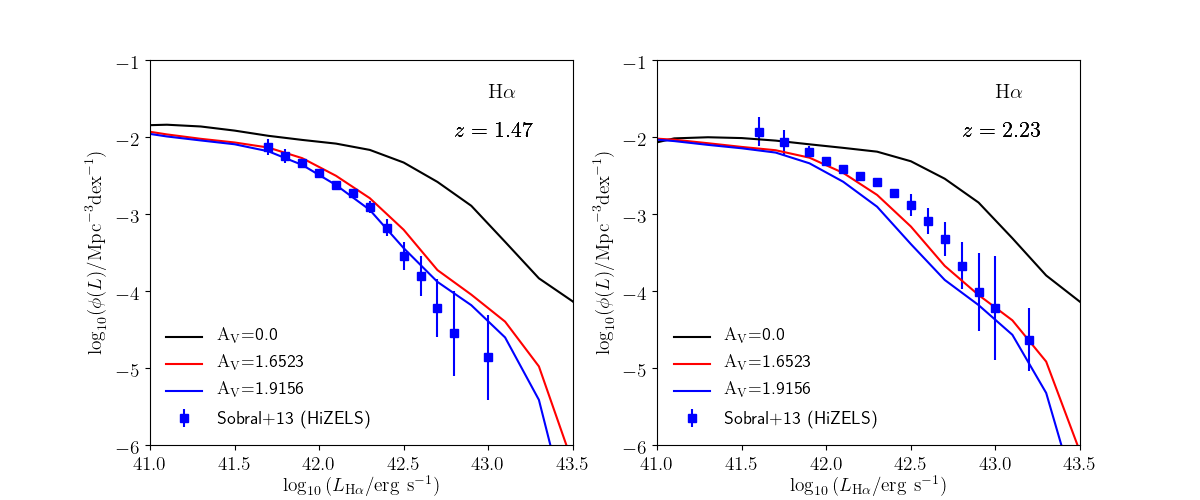}
\includegraphics[width=18.5cm, height=7cm]{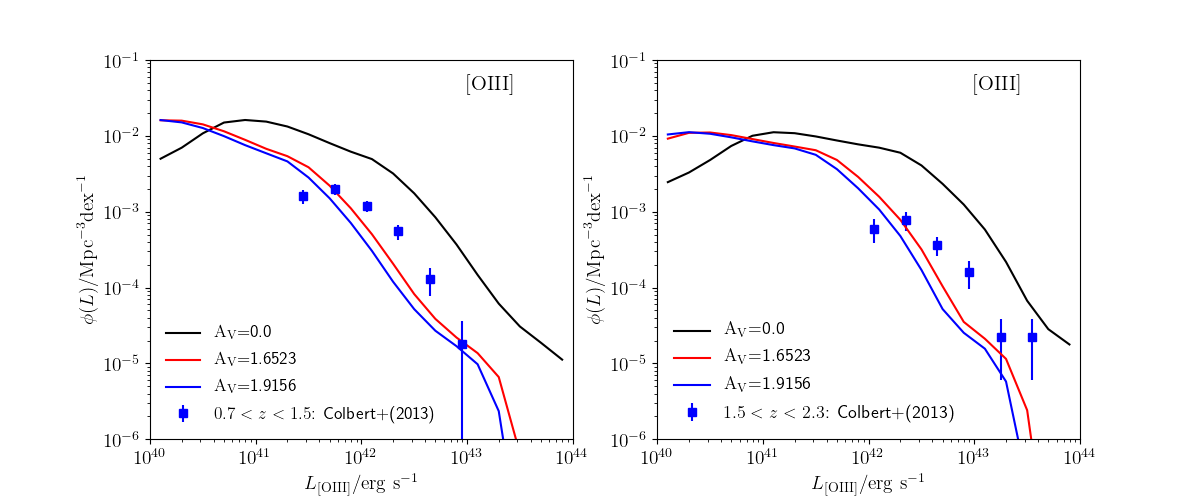}
\caption{\textbf{Top}: the prediction of the LF for H$\alpha$ (left) and [OIII] emission line galaxies for different redshifts and dust models. The dust free model ($\text{A}_{\text{V}}=0$) shows the intrinsic distribution, while $\text{A}_{\text{V}}=1.652$ and $\text{A}_{\text{V}}=1.916$ shows two dust models calibrated to match WISP number counts and HiZELS H$\alpha$ LF measurements, respectively. \textbf{Middle}: The comparison of the H$\alpha$ LF with HiZELS at two high redshifts. The dust model with $\text{A}_{\text{V}}=1.916$ can match the observation at $z=1.47$ perfectly since this measurement is used in the calibration of Galacticus. \textbf{Bottom}: The LF prediction of [OIII] ELGs compared with observational data from \citet{Colbert_2013}. Note that these measurements are not used in the calibration, but the amplitude is roughly consistent.}
\label{fig:LF}
\end{center}
\end{figure*}

In the top row of Figure \ref{fig:LF}, we present the luminosity function (LF) of H$\alpha$ and [OIII] emission line galaxies at $1<z<3$, with both dust free and dust attenuated results. The evolution with redshift indicates the star forming history of the galaxies. In order to validate the simulation, we compare the LF of H$\alpha$ and [OIII] emission lines with current observations at selected redshifts. The middle row shows the LF of H$\alpha$ galaxies compared with HiZELS measurements, which was used to calibrate the SAM model. In particular, the dust model with $A_{V}=1.916$ is chosen to match the LF at $z=1.47$ from HiZELS. Our model underestimates the LF at higher redshift, indicating either the need for improvement in the Galacticus model, or that a redshift dependent dust model is required. The bottom panel shows the comparison of [OIII] emission line galaxies with WISP measurements. Our model shows mild deviation compared with WISP measurements, however the amplitude is roughly consistent. The performance can be improved by calibrating the SAM with more observational data sets. 

\subsection{Sample selection for the Roman galaxy redshift survey}

\begin{figure}
\begin{center}
\includegraphics[width=8.5cm]{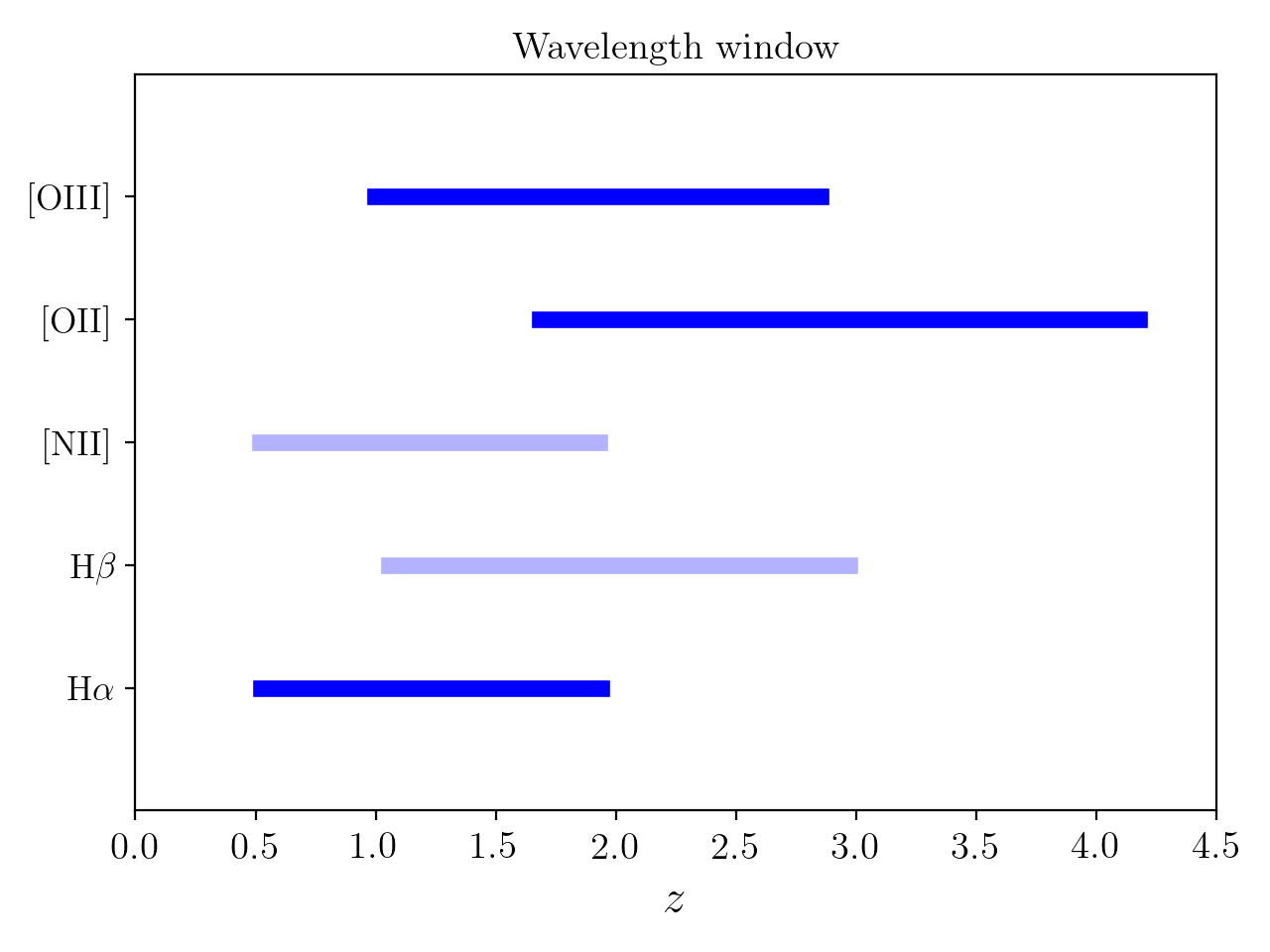}
\caption{The redshift coverage of the main emission lines of Roman HLSS and their main contaminant, determined by the wavelength range of Roman grism which is $1\sim1.93$ micron.}
\label{fig:redshift_range}
\end{center}
\end{figure}

In this paper, we focus on the forecast of galaxy linear bias for Roman HLSS, but the results can also be applicable to surveys like the one planned for Euclid. The observing strategies can impact the galaxy selection and thus linear bias measurements. Roman grism has a wavelength range of $1.0-1.93$ microns, which determines the redshift range for the emission lines of interest, as shown In Figure \ref{fig:redshift_range}, which includes the three primary lines H$\alpha$, [OII] and [OIII]. We also shows the [NII] and H$\beta$ lines as they are the main contaminants to H$\alpha$ and [OIII] respectively, due to the closeness of the emission line wavelength. Since the current Galacticus SAM model significantly underestimates the strength of [OII] emission, we will only consider H$\alpha$ and [OIII] lines throughout this paper, as they define the expected Roman galaxy samples.

One of the key characteristics of a survey is its depth, or sensitivity, i.e. the emission line flux limit for an ELG survey. We consider three different line flux limits in our analysis: $0.5\times10^{-17}\mathrm{erg}/\mathrm{s}/\mathrm{cm}^{2}$ as a reference case,  $1.0\times10^{-16}\mathrm{erg}/\mathrm{s}/\mathrm{cm}^{2}$ as the $6.5\sigma$ nominal depth for Roman HLSS, and $2.0\times10^{-16}\mathrm{erg}/\mathrm{s}/\mathrm{cm}^{2}$ as the $3.5\sigma$ depth of a Euclid-like GRS. The faint limit of $0.5\times10^{-17}\mathrm{erg}/\mathrm{s}/\mathrm{cm}^{2}$ is included to facilitate a depth versus area optimization study of the Roman HLSS. As the primary target, H$\alpha$ emission line is detectable at $z<2.0$. Therefore we split the analysis into two subsamples with $z>2.0$ and $z<2.0$.

At $z<2.0$, Roman science requirements specify that at least two emission lines are used in measuring a redshift, with the strong line above the line flux limit. The detection of the second line may not require its strength to be above the line flux limit at 6.5$\sigma$, thus we allow for different thresholds for the second emission line.
At $z>2.0$, we only consider the [OIII] line, required to be observed above the flux limit at 6.5$\sigma$, since we are not including [OII] in this study due to the current limits of Galacticus. The impact of the line flux threshold of the [OII] line will be studied in future work.

For the galaxies with $z<2.0$, we first compute the emission line flux for H$\alpha$ and [OIII], then we set the flux limits by two variables. The first is $f_{\mathrm{lim,1}}$, this is the lower limit of the stronger line (either H$\alpha$ or [OIII]), as chosen above. The second is $f_{\mathrm{lim, 2R}}$ which sets the lower limit of the weak line in units of $f_{\mathrm{lim,1}}$, therefore $f_{\mathrm{lim, 2R}}$ is dimensionless. We choose three values and investigate the impact on the large scale structure analysis: $f_{\mathrm{lim, 2R}}=0.25, 0.5, 1.0$. For instance when $f_{\mathrm{lim,1}}=1.0\times10^{-16}\mathrm{erg}/\mathrm{s}/\mathrm{cm}^{2}$ and $f_{\mathrm{lim, 2R}}=0.5$, we select galaxies with the strongest emission line brighter than $f_{\mathrm{lim,1}}$, and the second emission line brighter than $f_{\mathrm{lim,1}}*f_{\mathrm{lim, 2R}}=0.5\times10^{-16}\mathrm{erg}/\mathrm{s}/\mathrm{cm}^{2}$. At $z>2.0$, we just apply $f_{\mathrm{lim,1}}$ to select [OIII] emitting galaxies. 

\begin{figure*}
\begin{center}
\includegraphics[width=18.5cm]{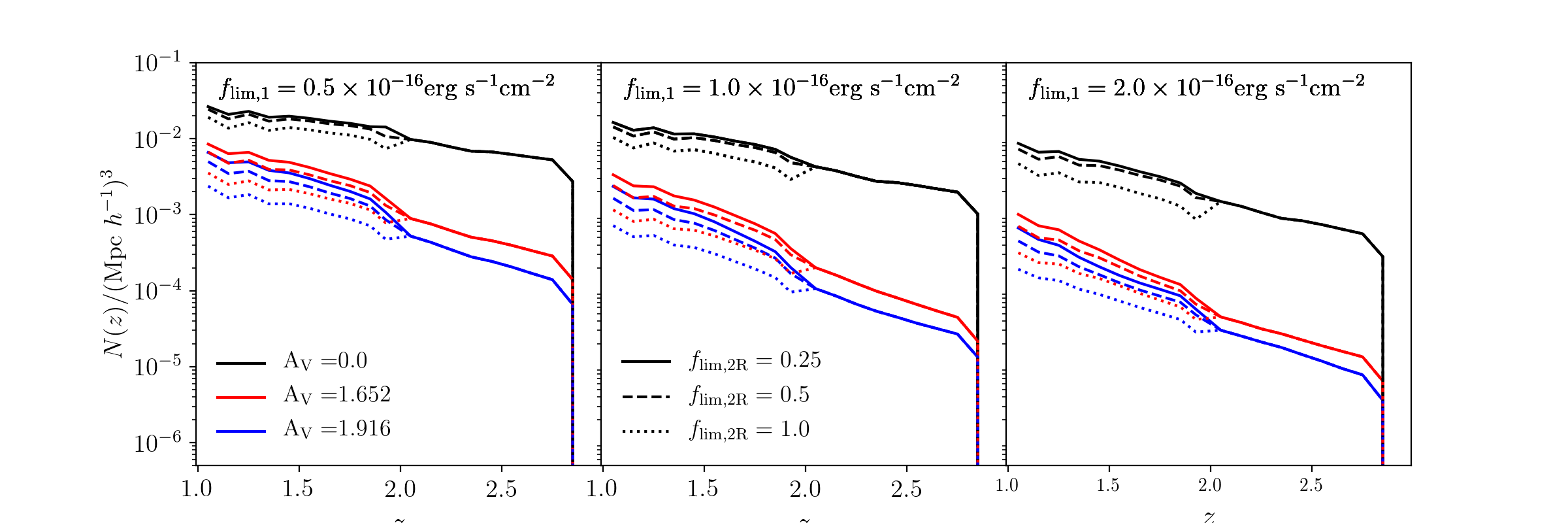}
\caption{The galaxy number density as a function of redshift, for different dust models and flux cuts. The curves of the same color merge at $z\sim2$ since there is only one flux cut for galaxies $z>2$. Three panels correspond to three values of $f_{\text{lim,1}}$, the flux limit of the strongest emission lines, the colors stand for different dust attenuation and the different line types denote the various values of $f_{\text{lim, 2R}}$. }
\label{fig:number_density}
\end{center}
\end{figure*}

In Figure \ref{fig:number_density}, we show the number densities of the selected galaxies with different flux limits. The curves of the same color merge at $z\sim2$ due to the selection algorithm since there is only one flux cut on [OIII]. The result shows a monotonic decrease as redshift increases. The flux limit parameter $f_{\mathrm{lim, 2R}}$ for the second emission line has weaker impact than $f_{\mathrm{lim,1}}$, but can be important when  brighter flux cut and stronger dust-attenuation are assumed. 

\begin{figure*}
\begin{center}
\includegraphics[width=18.5cm]{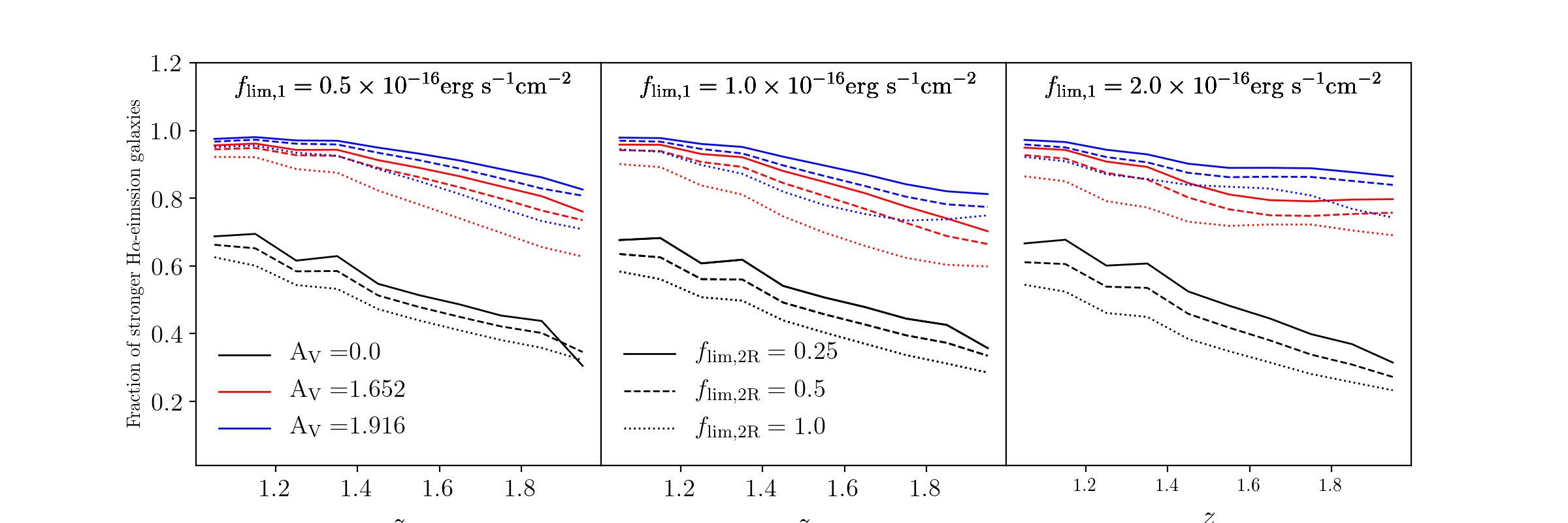}
\caption{The fraction of galaxies with H$\alpha$ as the strongest emission line, with dependence on dust model and flux cuts. This result is only for samples with $z<2.0$ since no H$\alpha$ emission is expected at higher redshift due to the grism wavelength coverage. Three panels correspond to three values of $f_{\text{lim, 1}}$, colors and line types have the same format as in Figure \ref{fig:number_density}.}
\label{fig:strongline_fraction}
\end{center}
\end{figure*}

For galaxies with $z<2.0$, both H$\alpha$ and [OIII] lines can be observed, and their relative strength may not be a constant. In Figure \ref{fig:strongline_fraction}, we plot the fraction of galaxies with H$\alpha$ as the stronger line, as a function of redshift. The left and middle panels show that as we go to higher redshifts, the H$\alpha$ dominance decreases with redshift regardless of the dust model, for sufficiently faint H$\alpha$ line flux cut. 
The right panel shows that this is not true for brighter H$\alpha$ line flux cut, for which the H$\alpha$ dominance flattens at higher redshifts. 
This indicates that there are more bright H$\alpha$ emitters than bright [OIII] emitters at high redshifts. In addition, Figure \ref{fig:strongline_fraction} shows the significant impact of the dust model. The intrinsic result (dust free with $A_{\mathrm{V}}=0$) shows that the H$\alpha$ dominance is around 30 to 70\% in this redshift range. However, when dust model is applied, almost all galaxies have stronger H$\alpha$ than [OIII] emission (>80\%), indicating that the [OIII] emission line experiences more dust attenuation as predicted by the Calzetti model. Since the H$\alpha$ emission is only present at $z<2$, we will refer to H$\alpha$ galaxies and galaxies at $z<2$ interchangeably in the following section, and similarly [OIII] galaxies refer to galaxies with $z>2$.

Because galaxies have peculiar velocities, the observed redshift is different from the cosmological redshift due to cosmic expansion. This RSD effect can change the measured galaxy distribution and the resultant clustering signal. In our simulation, we add this effect into the galaxy catalog by perturbing the cosmological redshift with $v_{p}/(ac)$, where $v_{p}$ is the line-of-sight component of the velocity, $a$ is the scale factor and $c$ is the speed of light. We will present clustering measurements in both real and redshift space in the following sections.

\section{Results}

\subsection{Galaxy clustering of H$\alpha$ galaxies} \label{sec:Halpha_galaxy}

Galaxies do not perfectly trace the underlying matter distribution. They preferentially live in the peaks of the matter density field. This makes the galaxies biased tracers of large scale structure, which preferentially sample the over-dense regions (\citealt{Kaiser_1984, Bardeen_1986}). In addition, the processes of galaxy formation can introduce additional deviations of the galaxy distribution from matter distribution. These factors result a relationship between the spatial distribution of galaxies and the dark matter density field: the galaxy bias. Neglecting the stochasticity and non-locality, the galaxy density contrast can be written as a function of the underlying dark matter density contrast on some scale (\citealt{Coil_2013}) $\delta_{g}=f(\delta_{m})$, where $\delta\equiv\rho/\bar{\rho}-1$ and $\bar{\rho}$ is the mean density. 

On large scales (also known as linear scales), where the density fluctuations are small and evolve linearly, we can expand the function $f$ and define the linear galaxy bias through $\delta_{g}=b\delta_{m}$. In terms of the 2-point correlation function, we can measure the galaxy bias by comparing the clustering amplitudes of galaxies and matter
\begin{equation}
    \xi_{gg}(r)=b^{2}(r)\,\xi_{mm}(r),
\end{equation}
where $\xi_{gg}$ and $\xi_{mm}$ are galaxy and matter correlation function as a function of spatial separation, respectively. 

With the simulated galaxy catalog from Galacticus, we compute the galaxy correlation function using the \cite{LS_1993} estimator,
\begin{equation}
    \xi_{r} = \frac{DD-2DR+RR}{RR},
\end{equation}
where $DD, DR$ and $RR$ are suitably normalized numbers of (weighted) data-data, data-random, and random-random pairs in each galaxy separation bin. The random catalogs are first generated with uniform distribution on a sphere and then truncated to have the same right ascension and declination boundary as the galaxy catalog. The redshifts of the random catalog are randomly drawn from the galaxy catalog to have the same radial distribution. The total number of randoms is 10 times larger than galaxy catalog to assure stable measurement of clustering. Following the same strategy as \cite{Merson_2019}, we measure the correlation function for each galaxy sample 5 times with different random catalogs. We measure the correlation function at spatial scales up to 150 Mpc$h^{-1}$. In Figure \ref{fig:Halpha_clustering}, we present the clustering measurement at a few redshift bins for galaxies with $1<z<2$ for both real and redshift space. We find that given our sample selection and dust model, the BAO peak can be recovered successfully. The worst case corresponds to high redshift galaxies with the brightest flux cut for the emission lines, which results in a low galaxy number density and thus the clustering signal is impacted by noise significantly. The attenuation from two dust models give similar impacts on the clustering amplitude, although their predictions for the number densities of ELGs are different (\citealt{Zhai_2019MNRAS}). The clustering amplitude in redshift space is higher than real space, consistent with expectation of the enhancement due to RSD effect. 

\subsection{Linear bias of H$\alpha$ galaxies} \label{sec:bias_Halpha}

\begin{figure*}
\begin{center}
\includegraphics[width=8.5cm]{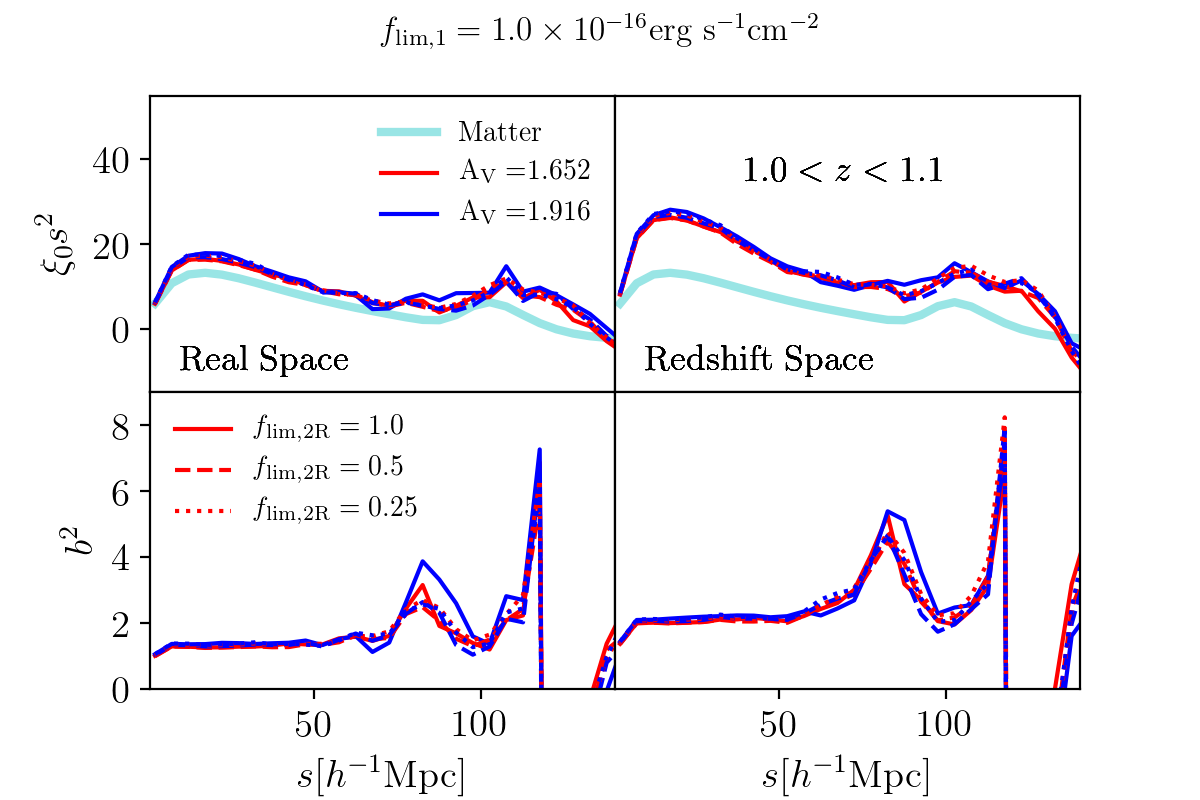}
\includegraphics[width=8.5cm]{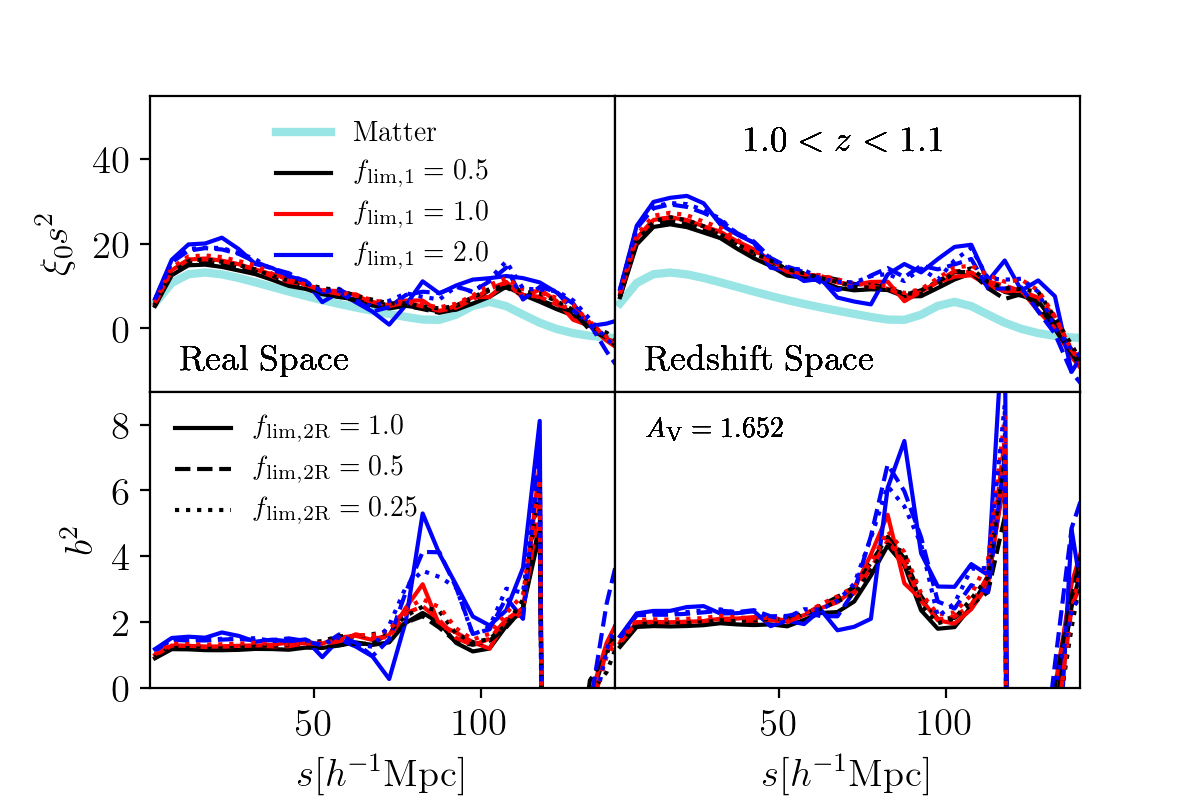}
\includegraphics[width=8.5cm]{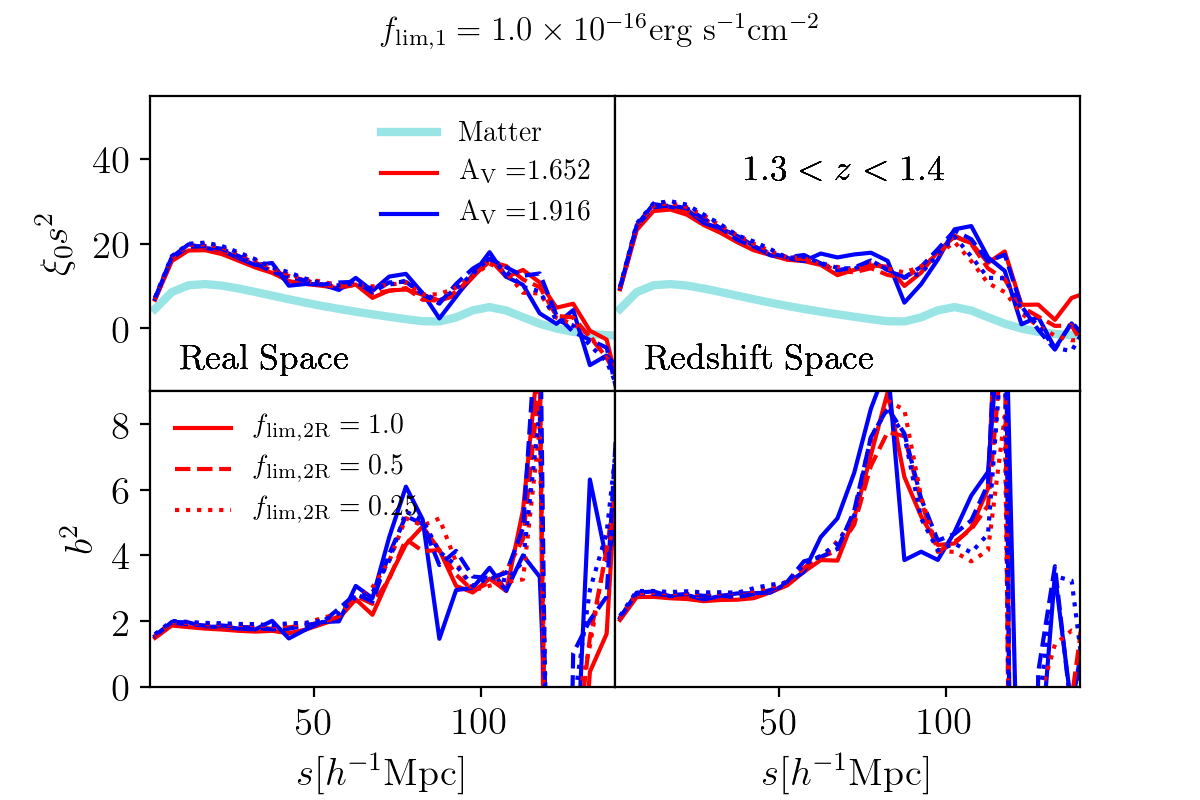}
\includegraphics[width=8.5cm]{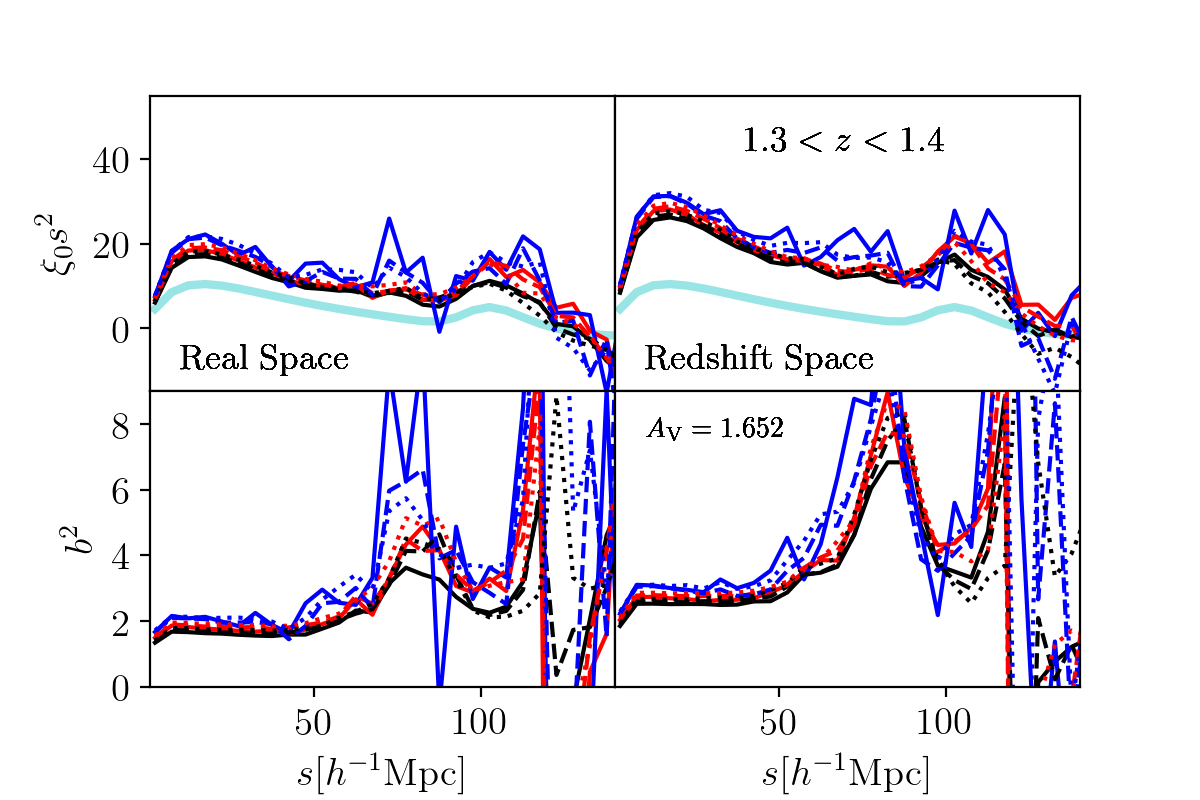}
\includegraphics[width=8.5cm]{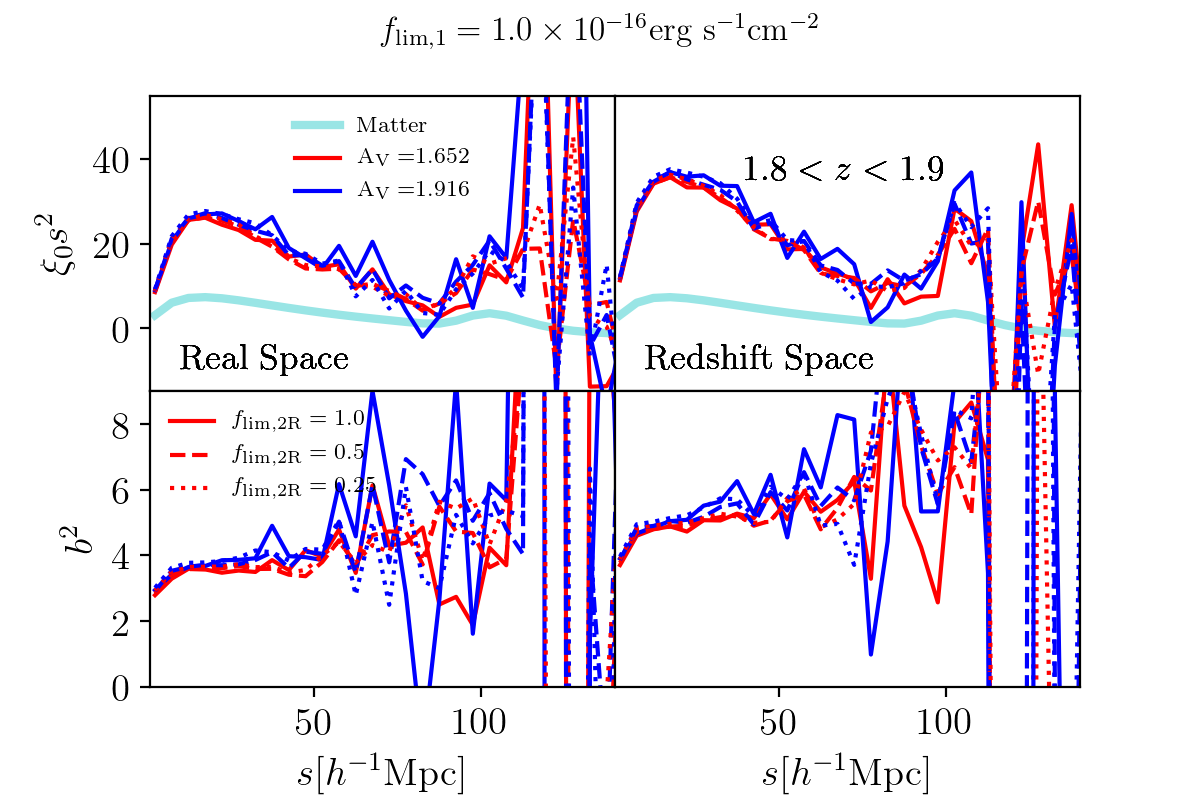}
\includegraphics[width=8.5cm]{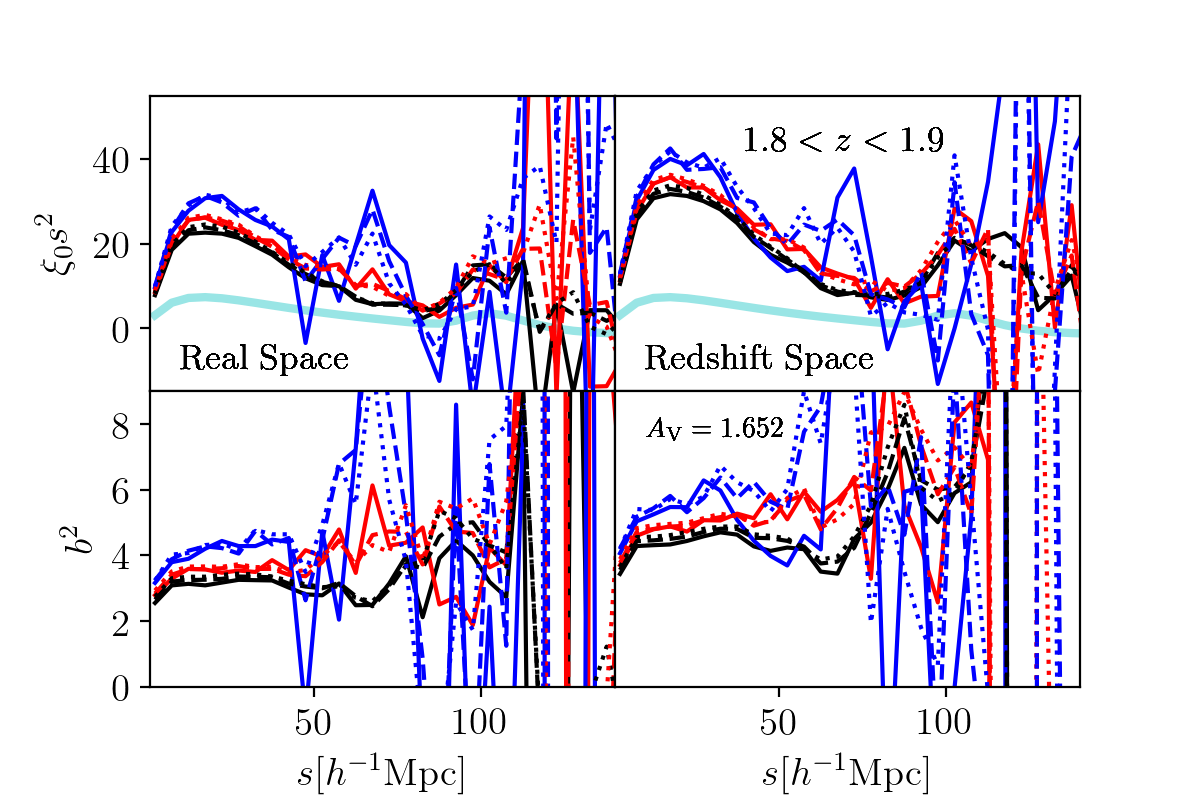}
\caption{The measurement of galaxy correlation function and bias for a few redshift slices, as indicated in each figure. The left column shows galaxies selected with $f_{\text{lim, 1}}=1.0\times10^{-16}\mathrm{erg}/\mathrm{s}/\mathrm{cm}^{2}$, and the line colors and styles indicate the effect of $f_{\text{lim, 2R}}$ and dust models respectively. The right column shows results with $A_{\text{V}}=1.652$, the line colors and styles indicate the effect of $f_{\text{lim, 1}}$ and $f_{\text{lim, 2R}}$ respectively. Note that to avoid cluttering, most of the figure legends are omitted in the middle and lower right panels for galaxies at $1.3<z<1.4$ and $1.8<z<1.9$; the line colors and styles are the same as the figures in the top right panel. The matter correlation function is shown as the thick cyan curve. The estimate of the galaxy bias $b^{2}$ is obtained by taking the ratio of the two point correlation function of galaxies and that of matter. Both real space and redshift space are shown for comparison.}
\label{fig:Halpha_clustering}
\end{center}
\end{figure*}

\begin{figure*}
\begin{center}
\includegraphics[width=18.5cm]{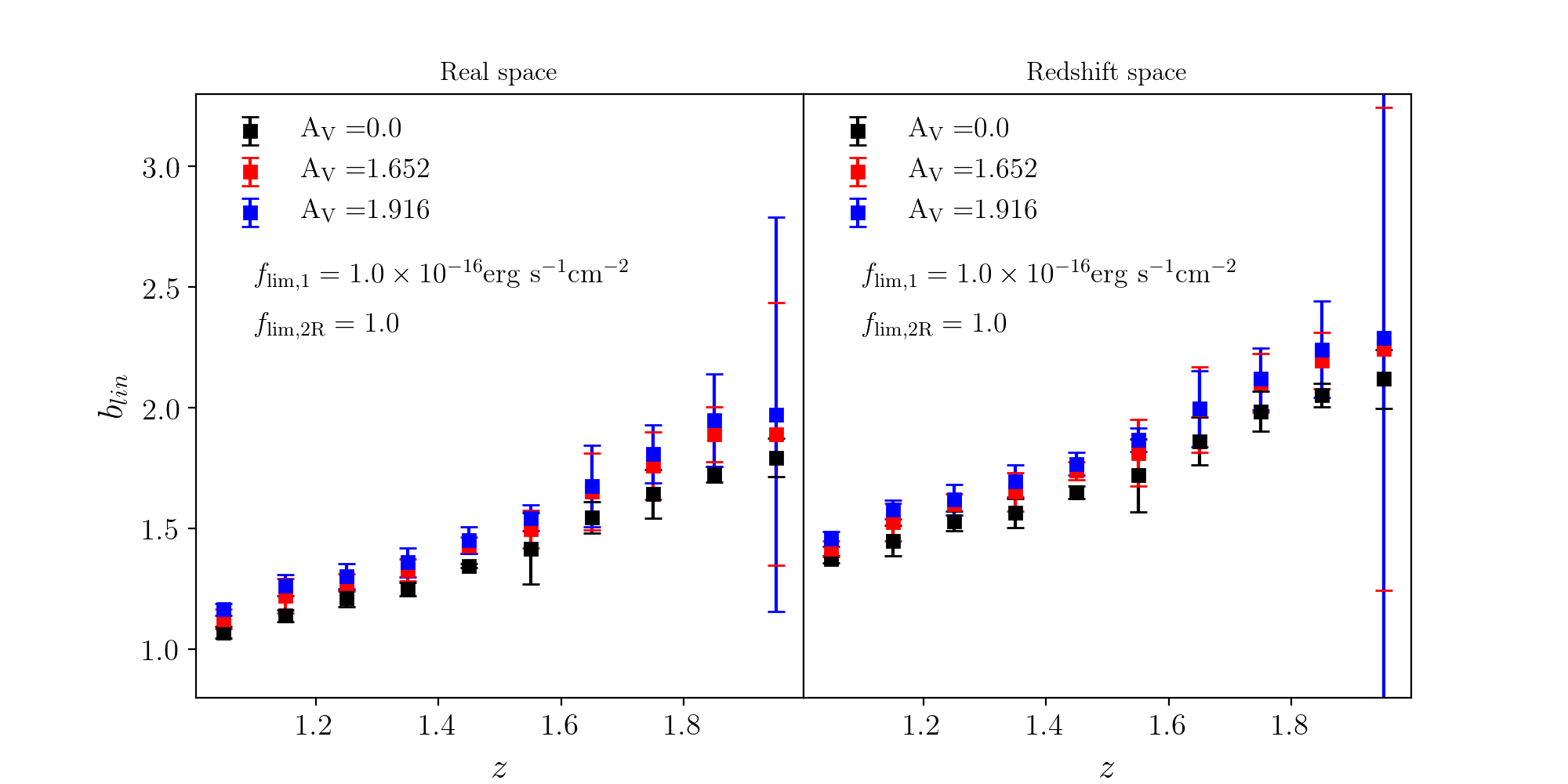}
\caption{Linear bias estimated as a function of redshift for both real and redshift space. The galaxies are chosen with $f_{\text{lim, 1}}=1.0\times10^{-16}\mathrm{erg}/\mathrm{s}/\mathrm{cm}^{2}$ and $f_{\text{lim, 2R}}=1.0$, i.e. only galaxies with both H$\alpha$ and [OIII] emission lines brighter than $1.0\times10^{-16}\mathrm{erg}/\mathrm{s}/\mathrm{cm}^{2}$ are selected. The dust free result (black square with error bar) is also shown for comparison with the two choices of $A_V$. $\mathrm{A}_{\mathrm{V}}=1.652$ corresponds to the calibration with WISP number counts, while $\mathrm{A}_{\mathrm{V}}=1.916$ is for the calibration based on H$\alpha$ luminosity function observed in HiZELS. The results with other selection criteria for the flux limit of emission lines are similar. We note that the linear bias measurement is close to a linear relation with respect to redshift.}
\label{fig:Halpha_bias}
\end{center}
\end{figure*}

\begin{figure*}
\begin{center}
\includegraphics[width=18.5cm]{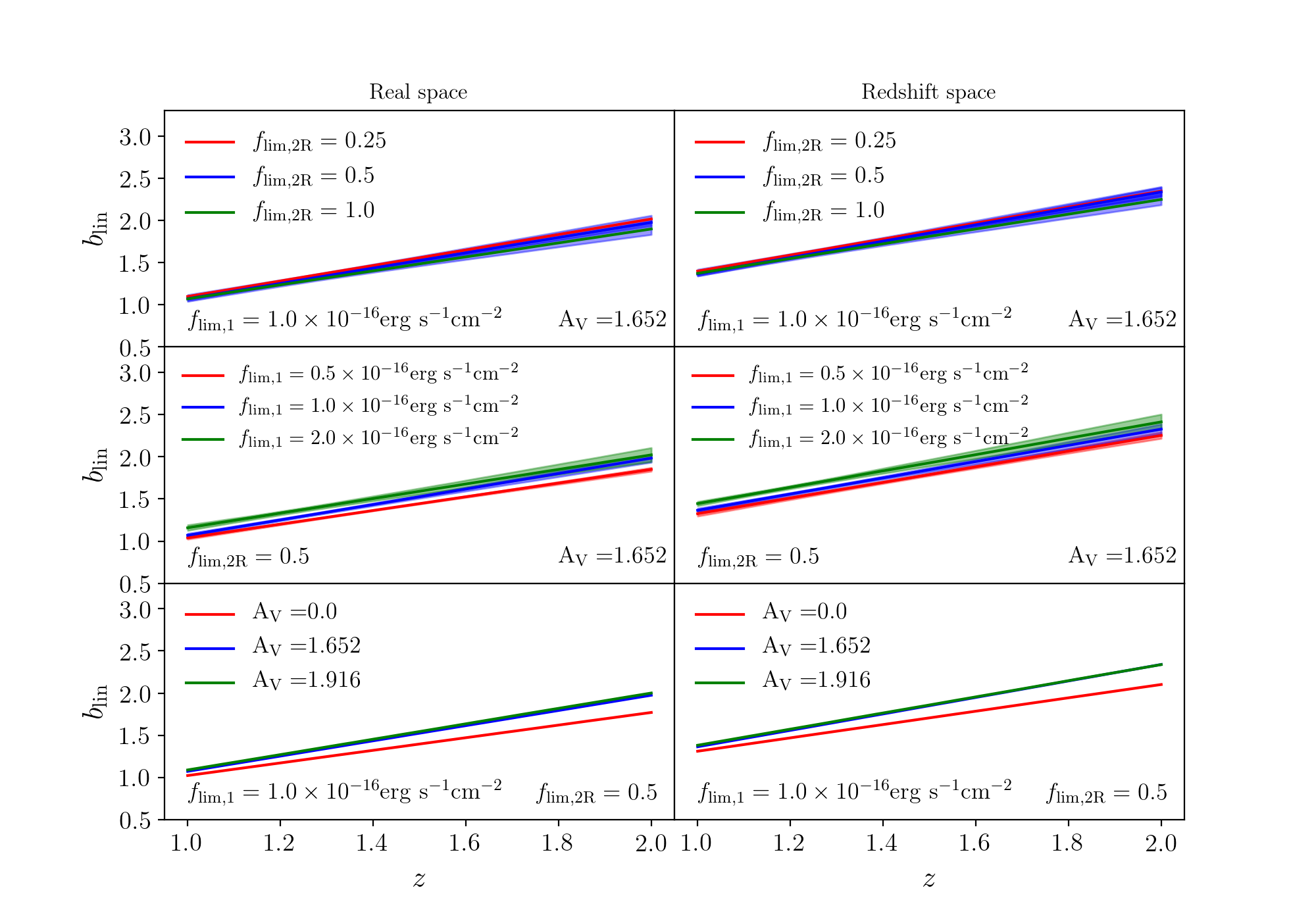}
\caption{Linear bias of H$\alpha$ galaxies as a linear function of redshift using clustering measurement within scales of $10<s<50h^{-1}$ Mpc. Both real (left) and redshift (right) space results are shown. The figure also displays the effect of changing the sample selection and dust attenuation: \textbf{Top:} impact of changing parameter $f_{\text{lim, 2R}}$, i.e. limit of the strength of the weak emission line. Galaxies are selected with the 6.5$\sigma$ nominal depth for the strongest emission line and the dust model can match WISP number counts. \textbf{Middle:} impact of the limit of the strongest emission line. The weak line has a limit of 50\% of $f_{\text{lim, 1}}$ and the dust model can match WISP number counts. \textbf{Bottom:} the effect of the dust model. The strongest emission line is brighter than the 6.5$\sigma$ nominal depth and the weaker line is brighter than $0.5\times10^{-16}\mathrm{erg}/\mathrm{s}/\mathrm{cm}^{2}$. The flux limit doesn't have a significant impact on the resulting bias measurement and reveals a weak tendency that higher flux limit preferentially selects more clustered galaxies, which is consistent with expectation. The dust model has a direct impact on the linear bias, but our calibrations based on WISP and HiZELS give consistent result. The shaded area represents inner 68\% distribution based on a MCMC test.}
\label{fig:Halpha_bias_fit}
\end{center}
\end{figure*}

In order to measure the linear bias, we compute the non-linear correlation function of dark matter $\xi_{mm}$ using the CLASS and Halofit functionality in the code Nbodykit (\citealt{Hand_2018}) with Planck 2016 cosmology, to be consistent with the dark matter simulation used in our SAM simulation. The result is shown as the cyan curve in each panel of Figure \ref{fig:Halpha_clustering}. The galaxy bias is computed by taking the ratio between the correlation functions of galaxies and dark matter. The lower row of each panel shows the resultant $b^{2}$. Same as for the galaxy correlation function, the bias is also obtained by using the mean of five repeat measurements. The error bars are omitted to clearly present the result. 

The figure shows that the bias is close to a constant at scales from 10 to 50 or 60 $h^{-1}$Mpc. At scales below 10 $h^{-1}$Mpc, the non-linearity of the dark matter dynamics comes into play and can induce scale dependent bias. On the other hand, at scales above 50 or 60 $h^{-1}$Mpc, the galaxy bias deviates from a constant value and presents complicated behavior, especially around the BAO scale. This feature is more significant in redshift space than in real space. This result is also pointed out in the earlier attempt presented in \cite{Merson_2019}, where the galaxies are H$\alpha$ only galaxies rather than the samples defined using two emission lines as in this paper. However, the cause for this distortion on the largest scales are similar, and a combination of several factors, such as the RSD effect, sample variance, mode coupling of the cosmic density perturbation and so on. 

With the measured galaxy correlation function, we can estimate the constant value of bias $b$ by fitting the $b^{2}$ as shown in Figure \ref{fig:Halpha_clustering} with a constant. Based on the measurement, we fit the data within $10<s<50$ $h^{-1}$Mpc. The uncertainty of the bias estimate adopts the same strategy as in \cite{Merson_2019} based on the root-mean-square (RMS) of the difference between $b$ and the fitted mean $b_{\text{lin}}$
\begin{equation}
    \delta b_{\text{lin}}=\sqrt{\frac{1}{N}\sum_{10<s<50}(b(s)-b_{\text{lin}})^2},
\end{equation}
where $N$ is the number of bins for the galaxy correlation function within $10<s<50$ $h^{-1}$Mpc. As an example, Figure \ref{fig:Halpha_bias} displays the resulting measurement of linear bias for galaxies selected with $f_{\text{lin}, 1}=1.0\times10^{-16}\mathrm{erg}/\mathrm{s}/\mathrm{cm}^{2}$ and $f_{\text{lin, 2R}}=1.0$, i.e. only galaxies with both H$\alpha$ and [OIII] flux brighter than $1.0\times10^{-16}\mathrm{erg}/\mathrm{s}/\mathrm{cm}^{2}$. It shows a close to linear relation for galaxy bias as a function of redshift, for both real and redshift space. The two dust models calibrated to match the WISP number counts and HiZELS H$\alpha$ LF respectively give quite consistent results. Note that the huge errorbar in the redshift space measurement for the galaxy subsample at highest redshift bin is due to the noisy measurement of the correlation function. 

The measured galaxy bias as a function of redshift can be simply described as $b_{\text{lin}}(z) = az+b$, where $a$ and $b$ are gradient and intercept respectively. Combining this model with the bias measurements, we construct a simple $\chi^{2}$ and perform a Monte Carlo Markov Chain (MCMC) test with the python code emcee (\citealt{Foreman-Mackey_2013}) to obtain the constraints on $a$ and $b$. We then sample from the posterior to estimate the 16 and 84 percentile as the uncertainty. In Figure \ref{fig:Halpha_bias_fit}, we show the fitting results for different dust models and sample selections. We can see that the dust model removes fainter and less massive galaxies. This can increase the average halo mass of the galaxy sample to increase the bias, as expected. The difference between the two dust models is not as significant as we find in the clustering measurements. The fitting result of the linear model is summarized in Table \ref{tab:Halpha_real} for real space and Table \ref{tab:Halpha_redshift} for redshift space with different sample selections and dust model. 

The flux limit of the strong line, i.e. $f_{\text{lin},1}$ has a direct impact on the linear bias. Increasing its value selects galaxies with brighter H$\alpha$ or [OIII] emissions. This is consistent with the relationship between H$\alpha$ or [OIII] luminosity and host halo mass (see e.g. \citealt{Zhai_2019MNRAS}). However, at higher redshifts, the impact is less significant, which is partially due to the larger fraction of [OIII] dominated galaxies and thus reduces the effect. The dependence of linear bias on $f_{\text{lin, 2R}}$, the flux limit of the secondary emission line is more complicated. The result doesn't present a monotonic relation. The reason is partially due to the flux ratio of H$\alpha$/[OIII], which has a clear dependence on the H$\alpha$ luminosity. However this dependence decreases with higher redshift (see for example Fig 7 in \citealt{Zhai_2019MNRAS}). Thus the scatter of flux ratio H$\alpha$/[OIII] at a given H$\alpha$ luminosity indicates that a galaxy with bright H$\alpha$ emission doesn't necessarily have bright [OIII] emission and vice versa. In general, we find that the dust models as well as the flux limit for the emission lines can affect our estimate of the linear bias at a few to $\sim$ten percent level.

\begin{table}
\centering
\begin{tabular}{llll}
\hline
$f_{\text{lin}, 1}=0.5$ &  $a$ & $b$   \\
\hline
$f_{\text{lin, 2R}}=0.25, A_{V}=0.0$  & $0.686\pm0.022$   & $0.349\pm0.034$ \\
$f_{\text{lin, 2R}}=0.25, A_{V}=1.652$  & $0.878\pm0.05$   & $0.166\pm0.069$ \\
$f_{\text{lin, 2R}}=0.25, A_{V}=1.915$  & $0.895\pm0.04$   & $0.156\pm0.056$ \\
\hline
$f_{\text{lin, 2R}}=0.5, A_{V}=0.0$  & $0.694\pm0.031$   & $0.31\pm0.045$ \\
$f_{\text{lin, 2R}}=0.5, A_{V}=1.652$  & $0.811\pm0.036$   & $0.228\pm0.052$ \\
$f_{\text{lin, 2R}}=0.5, A_{V}=1.915$  & $0.866\pm0.031$   & $0.175\pm0.044$ \\
\hline
$f_{\text{lin, 2R}}=1.0, A_{V}=0.0$  & $0.785\pm0.028$   & $0.171\pm0.041$ \\
$f_{\text{lin, 2R}}=1.0, A_{V}=1.652$  & $0.834\pm0.046$   & $0.183\pm0.063$ \\
$f_{\text{lin, 2R}}=1.0, A_{V}=1.915$  & $0.862\pm0.049$   & $0.167\pm0.068$ \\
\hline
$f_{\text{lin}, 1}=1.0$ &  $a$ & $b$   \\
\hline
$f_{\text{lin, 2R}}=0.25, A_{V}=0.0$  & $0.806\pm0.038$   & $0.192\pm0.056$ \\
$f_{\text{lin, 2R}}=0.25, A_{V}=1.652$  & $0.918\pm0.055$   & $0.181\pm0.069$ \\
$f_{\text{lin, 2R}}=0.25, A_{V}=1.915$  & $0.916\pm0.063$   & $0.191\pm0.079$ \\
\hline
$f_{\text{lin, 2R}}=0.5, A_{V}=0.0$  & $0.743\pm0.049$   & $0.281\pm0.063$ \\
$f_{\text{lin, 2R}}=0.5, A_{V}=1.652$  & $0.915\pm0.06$   & $0.156\pm0.078$ \\
$f_{\text{lin, 2R}}=0.5, A_{V}=1.915$  & $0.907\pm0.074$   & $0.182\pm0.095$ \\
\hline
$f_{\text{lin, 2R}}=1.0, A_{V}=0.0$  & $0.778\pm0.041$   & $0.226\pm0.058$ \\
$f_{\text{lin, 2R}}=1.0, A_{V}=1.652$  & $0.844\pm0.088$   & $0.221\pm0.118$ \\
$f_{\text{lin, 2R}}=1.0, A_{V}=1.915$  & $0.788\pm0.085$   & $0.337\pm0.106$ \\
\hline
$f_{\text{lin}, 1}=2.0$ &  $a$ & $b$   \\
\hline
$f_{\text{lin, 2R}}=0.25, A_{V}=0.0$  & $0.809\pm0.047$   & $0.247\pm0.066$ \\
$f_{\text{lin, 2R}}=0.25, A_{V}=1.652$  & $0.88\pm0.071$   & $0.283\pm0.095$ \\
$f_{\text{lin, 2R}}=0.25, A_{V}=1.915$  & $0.853\pm0.111$   & $0.357\pm0.129$ \\
\hline
$f_{\text{lin, 2R}}=0.5, A_{V}=0.0$  & $0.828\pm0.051$   & $0.208\pm0.07$ \\
$f_{\text{lin, 2R}}=0.5, A_{V}=1.652$  & $0.859\pm0.102$   & $0.297\pm0.127$ \\
$f_{\text{lin, 2R}}=0.5, A_{V}=1.915$  & $0.87\pm0.132$   & $0.317\pm0.158$ \\
\hline
$f_{\text{lin, 2R}}=1.0, A_{V}=0.0$  & $0.857\pm0.041$   & $0.168\pm0.058$ \\
$f_{\text{lin, 2R}}=1.0, A_{V}=1.652$  & $0.871\pm0.132$   & $0.31\pm0.178$ \\
$f_{\text{lin, 2R}}=1.0, A_{V}=1.915$  & $0.807\pm0.229$   & $0.428\pm0.287$ \\
\hline
\end{tabular}
\caption{The fitting result for the galaxy bias as a linear function of redshift: $b_{\text{lin}}(z)=az+b$, estimated from clustering measurement in real space for H$\alpha$ galaxies. The flux limit $f_{\text{lin}, 1}$ is in unit of $\times10^{-16}\mathrm{erg}/\mathrm{s}/\mathrm{cm}^{2}$, and parameter $f_{\text{lin, 2R}}$ is dimensionless.}
\label{tab:Halpha_real}
\end{table}

\begin{table}
\centering
\begin{tabular}{llll}
\hline
$f_{\text{lin}, 1}=0.5$ &  $a$ & $b$   \\
\hline
$f_{\text{lin, 2R}}=0.25, A_{V}=0.0$  & $0.727\pm0.046$   & $0.598\pm0.07$ \\
$f_{\text{lin, 2R}}=0.25, A_{V}=1.652$  & $0.886\pm0.054$   & $0.472\pm0.075$ \\
$f_{\text{lin, 2R}}=0.25, A_{V}=1.915$  & $0.943\pm0.054$   & $0.416\pm0.075$ \\
\hline
$f_{\text{lin, 2R}}=0.5, A_{V}=0.0$  & $0.699\pm0.04$   & $0.617\pm0.057$ \\
$f_{\text{lin, 2R}}=0.5, A_{V}=1.652$  & $0.929\pm0.062$   & $0.393\pm0.089$ \\
$f_{\text{lin, 2R}}=0.5, A_{V}=1.915$  & $0.908\pm0.036$   & $0.442\pm0.049$ \\
\hline
$f_{\text{lin, 2R}}=1.0, A_{V}=0.0$  & $0.799\pm0.045$   & $0.476\pm0.07$ \\
$f_{\text{lin, 2R}}=1.0, A_{V}=1.652$  & $0.888\pm0.051$   & $0.43\pm0.07$ \\
$f_{\text{lin, 2R}}=1.0, A_{V}=1.915$  & $0.891\pm0.061$   & $0.434\pm0.085$ \\
\hline
$f_{\text{lin}, 1}=1.0$ &  $a$ & $b$   \\
\hline
$f_{\text{lin, 2R}}=0.25, A_{V}=0.0$  & $0.807\pm0.051$   & $0.525\pm0.075$ \\
$f_{\text{lin, 2R}}=0.25, A_{V}=1.652$  & $0.938\pm0.065$   & $0.462\pm0.08$ \\
$f_{\text{lin, 2R}}=0.25, A_{V}=1.915$  & $0.942\pm0.082$   & $0.466\pm0.104$ \\
\hline
$f_{\text{lin, 2R}}=0.5, A_{V}=0.0$  & $0.793\pm0.041$   & $0.519\pm0.06$ \\
$f_{\text{lin, 2R}}=0.5, A_{V}=1.652$  & $0.962\pm0.075$   & $0.406\pm0.095$ \\
$f_{\text{lin, 2R}}=0.5, A_{V}=1.915$  & $0.944\pm0.075$   & $0.442\pm0.097$ \\
\hline
$f_{\text{lin, 2R}}=1.0, A_{V}=0.0$  & $0.804\pm0.048$   & $0.522\pm0.06$ \\
$f_{\text{lin, 2R}}=1.0, A_{V}=1.652$  & $0.881\pm0.094$   & $0.489\pm0.122$ \\
$f_{\text{lin, 2R}}=1.0, A_{V}=1.915$  & $0.826\pm0.087$   & $0.569\pm0.112$ \\
\hline
$f_{\text{lin}, 1}=2.0$ &  $a$ & $b$   \\
\hline
$f_{\text{lin, 2R}}=0.25, A_{V}=0.0$  & $0.868\pm0.054$   & $0.488\pm0.078$ \\
$f_{\text{lin, 2R}}=0.25, A_{V}=1.652$  & $0.965\pm0.103$   & $0.494\pm0.125$ \\
$f_{\text{lin, 2R}}=0.25, A_{V}=1.915$  & $0.922\pm0.105$   & $0.57\pm0.121$ \\
\hline
$f_{\text{lin, 2R}}=0.5, A_{V}=0.0$  & $0.85\pm0.062$   & $0.486\pm0.086$ \\
$f_{\text{lin, 2R}}=0.5, A_{V}=1.652$  & $0.979\pm0.096$   & $0.463\pm0.114$ \\
$f_{\text{lin, 2R}}=0.5, A_{V}=1.915$  & $0.946\pm0.144$   & $0.532\pm0.178$ \\
\hline
$f_{\text{lin, 2R}}=1.0, A_{V}=0.0$  & $0.934\pm0.042$   & $0.4\pm0.056$ \\
$f_{\text{lin, 2R}}=1.0, A_{V}=1.652$  & $0.948\pm0.153$   & $0.515\pm0.205$ \\
$f_{\text{lin, 2R}}=1.0, A_{V}=1.915$  & $0.893\pm0.207$   & $0.618\pm0.297$ \\
\hline
\end{tabular}
\caption{The same as Table \ref{tab:Halpha_real}, but for redshift space.}
\label{tab:Halpha_redshift}
\end{table}

\subsection{HOD of H$\alpha$ galaxies} \label{sec:HOD_halpha}

\begin{figure*}
\begin{center}
\includegraphics[width=18.5cm]{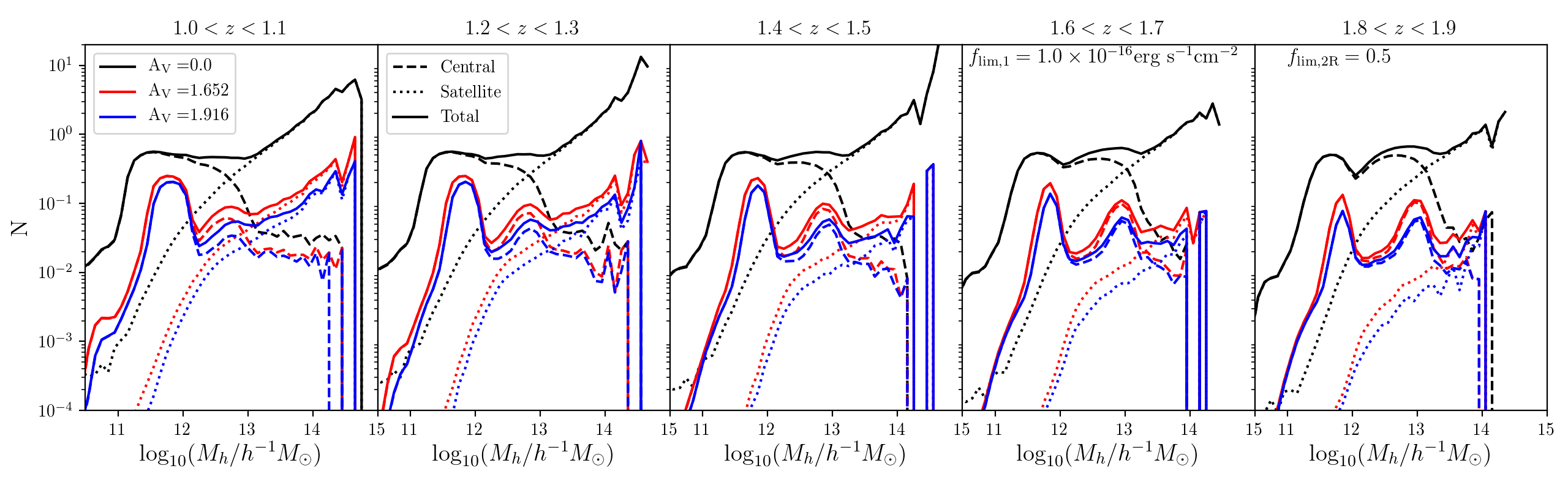}
\includegraphics[width=18.5cm]{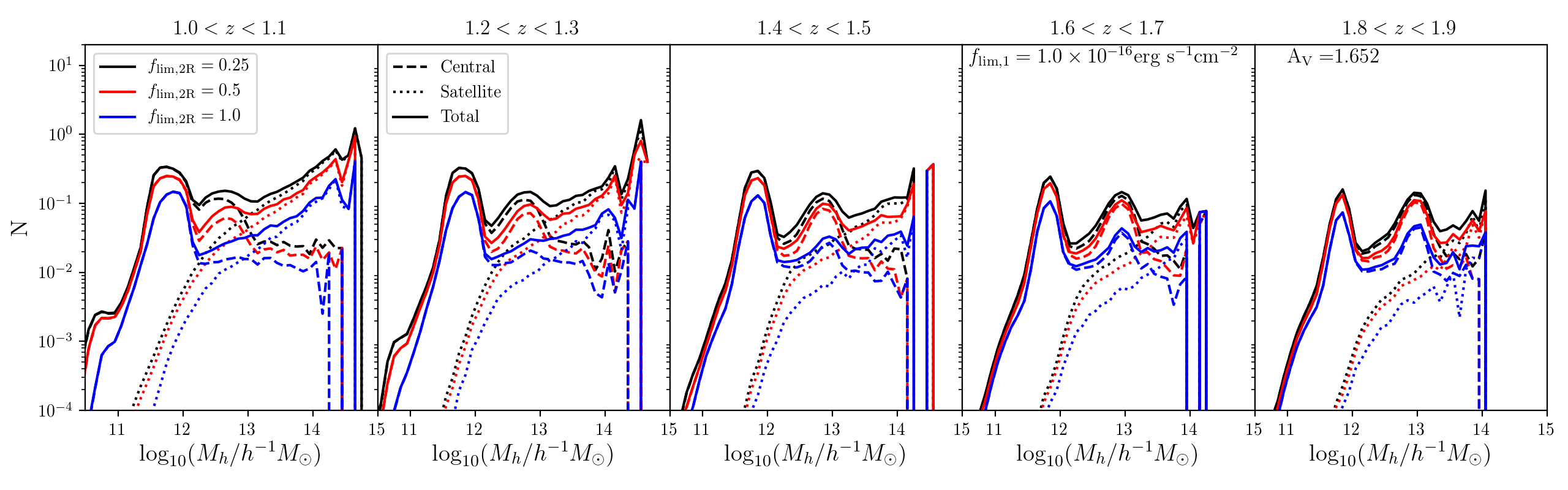}
\includegraphics[width=18.5cm]{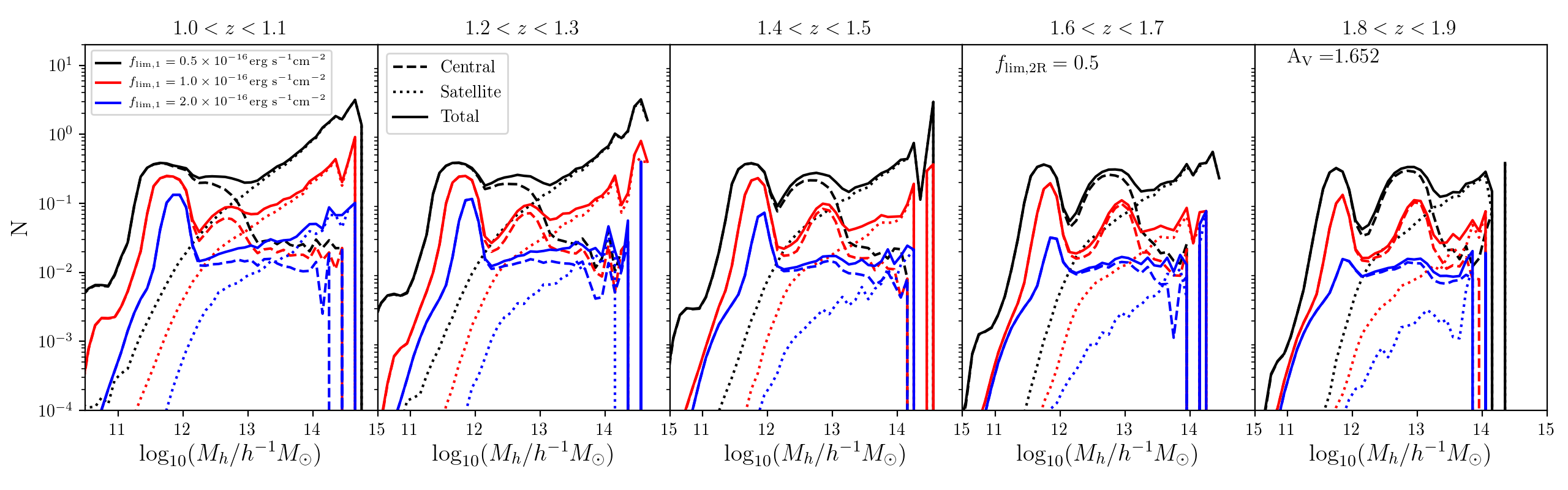}
\caption{Halo Occupation Distribution (HOD) of the H$\alpha$ galaxies for redshift slices as indicated in each panel. The effect of changing the selection parameters and dust model is also shown. \textbf{Top:} The effect of changing dust attenuation for galaxies selected with $f_{\text{lin}, 1}=1.0\times10^{-16}\mathrm{erg}/\mathrm{s}/\mathrm{cm}^{2}$ and $f_{\text{lin, 2R}}=0.5$, the intrinsic distribution without dust-attenuation ($\text{A}_{V}=0$) is also shown as black curves. \textbf{Middle:} The effect of selection parameter $f_{\text{lin, 2R}}$ for galaxies selected with $1.0\times10^{-16}\mathrm{erg}/\mathrm{s}/\mathrm{cm}^{2}$ for the strongest emission line and WISP-calibrated dust model. \textbf{Bottom:} The effect of changing the flux limit of the strongest line $f_{\text{lin}, 1}$ for galaxy samples defined with $f_{\text{lin, 2R}}=0.5$ and WISP-calibrated dust model. We present the occupation of centrals (dashed), satellites (dotted) and the total (solid) respectively. }
\label{fig:HODz1}
\end{center}
\end{figure*}

Halo Occupation Distribution (HOD) is a statistical approach to describe the connection between galaxies and dark matter halos. It has been used to interpret the observations of galaxy clustering over a wide range of redshifts and luminosities, see e.g. \cite{Zheng_2005, Zheng_2007, Zehavi_2011, CMASS_Martin, Zhai_2017}. In practice, it splits the galaxies into centrals and satellites. The investigations of massive galaxies have built simple parametrizations to describe the functional forms for the centrals and satellites. However, our understanding of the HOD of the ELGs is relatively poor, although some pioneering work has been done based on simulated or observed ELGs, for instance \cite{Geach_2012, Contreras_2013, Cochrane_2018, Gonzalez-Perez_2018, Avila_2020} and references therein. 

The SAM simulation in our work provides a reasonable framework for the measurement of HOD of the ELGs. In Figure \ref{fig:HODz1}, we present the measured HOD using different selections and dust models for galaxies in a few redshift slices. The first prominent feature is that the HOD for centrals has a double peak shape as a function of halo mass. The valley is around $10^{12.4}M_{\sun}h^{-1}$ and this position has no significant dependence on sample selection. This double peak HOD for centrals differs from the results of studies based on luminous red galaxies (LRGs), which are consistent with a monotonic function for central occupation. We note that this double peak behavior in the HOD can be a combination of different factors, including sample selection, dust model and galaxy formation physics. In the earlier attempt of \cite{Merson_2019}, the HOD for H$\alpha$ galaxies is measured based on Millennium simulation. Regardless of the different sample selections and parameter sets for the SAM, they also find similar, but weaker, behavior as ours. This feature becomes less pronounced when a higher threshold for luminosity is adopted, consistent with the tendency shown in Figure \ref{fig:HODz1}. When we increase the flux limit for the emission line or dust attenuation to reduce the galaxy number density, the double peak feature becomes less significant. This indicates that the occupancy of the second peak is dominated by less mass galaxies. Previous work based on observational data or simulation shows that the occupation of central galaxies only peaks at low mass range and quickly declines at high mass end (\citealt{Contreras_2013, Gonzalez-Perez_2018, Avila_2020, Hadzhiyska_2020}). 

The double-peaked nature present in the predicted HODs can be traced back to the presence of two distinct sequences in the plane of cenral galaxy star formation rate and halo mass in the Galactics models. We have examined the origin of these two sequences and find that they are the result of the fact that, in the UNIT simulation merger trees, some significant fraction of dark matter halos undergo periods of mass loss (i.e. their total mass decreases with time).

Galacticus assumes that halos accrete baryons from their surroundings at a rate proportional to their halo mass growth rate. During periods of mass loss in a halo, Galacticus instead holds the baryonic content of the halo fixed, and let it begin to increase again when the halo has grown beyond its previous greatest mass.

As a consequence, halos undergoing mass loss have no new gas supply, and so the star formation rate of their central galaxies quickly declines, leading to the formation of a second sequence of galaxies in the plane of star formation rate and halo mass.

These periods of mass loss from halos may be physical (driven by merging events which cause mass to be ejected), or may be purely numerical in origin (due to the choice of halo mass definition, or to failings in the halo finder and merger tree builder to link halos together over time). A detailed examination of the origins of these periods of mass loss, and how best to model their effect on the baryonic content of halos, is beyond the scope of this paper, but will be explored in a future work.

The satellite occupation from Galacticus is consistent with expectations, and can be represented by a functional form close to a power law, which is similar to massive galaxies at lower redshift. However, we note that this power low can break for high redshift galaxies.

\subsection{[OIII] galaxies} \label{sec:OIII}

\begin{figure*}
\begin{center}
\includegraphics[width=8.5cm]{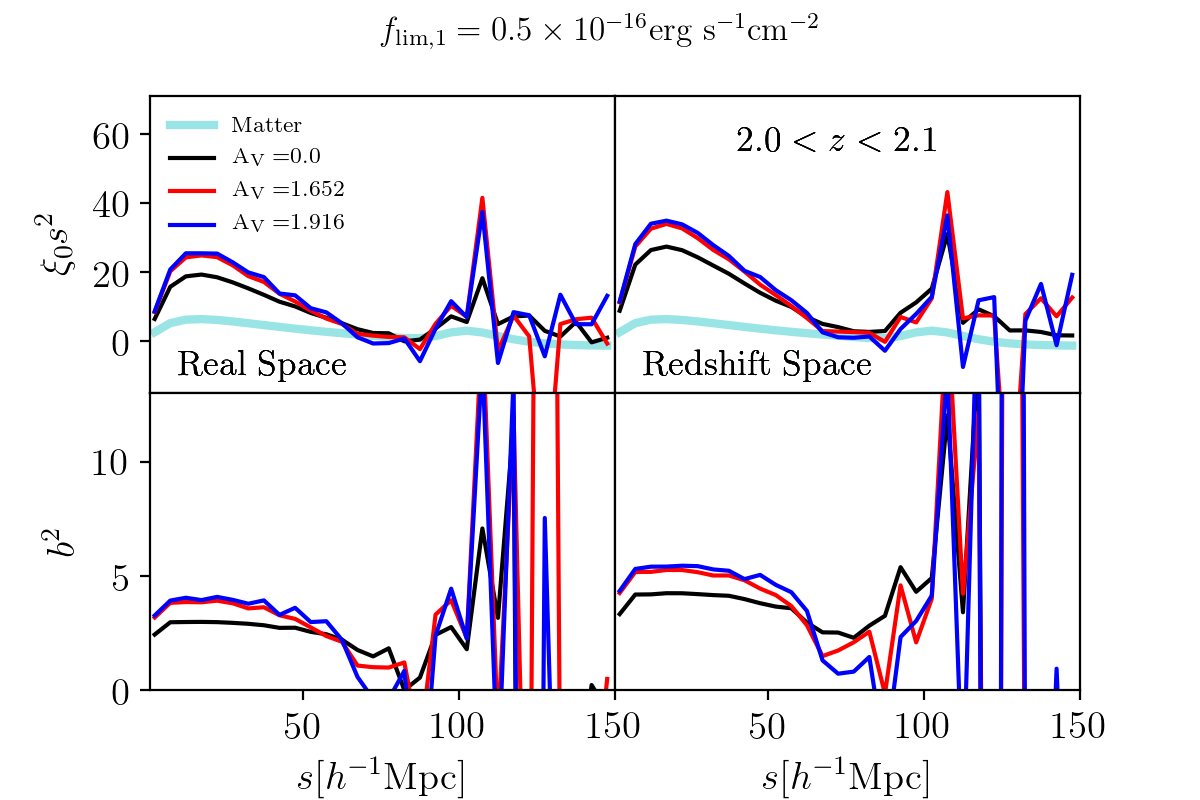}
\includegraphics[width=8.5cm]{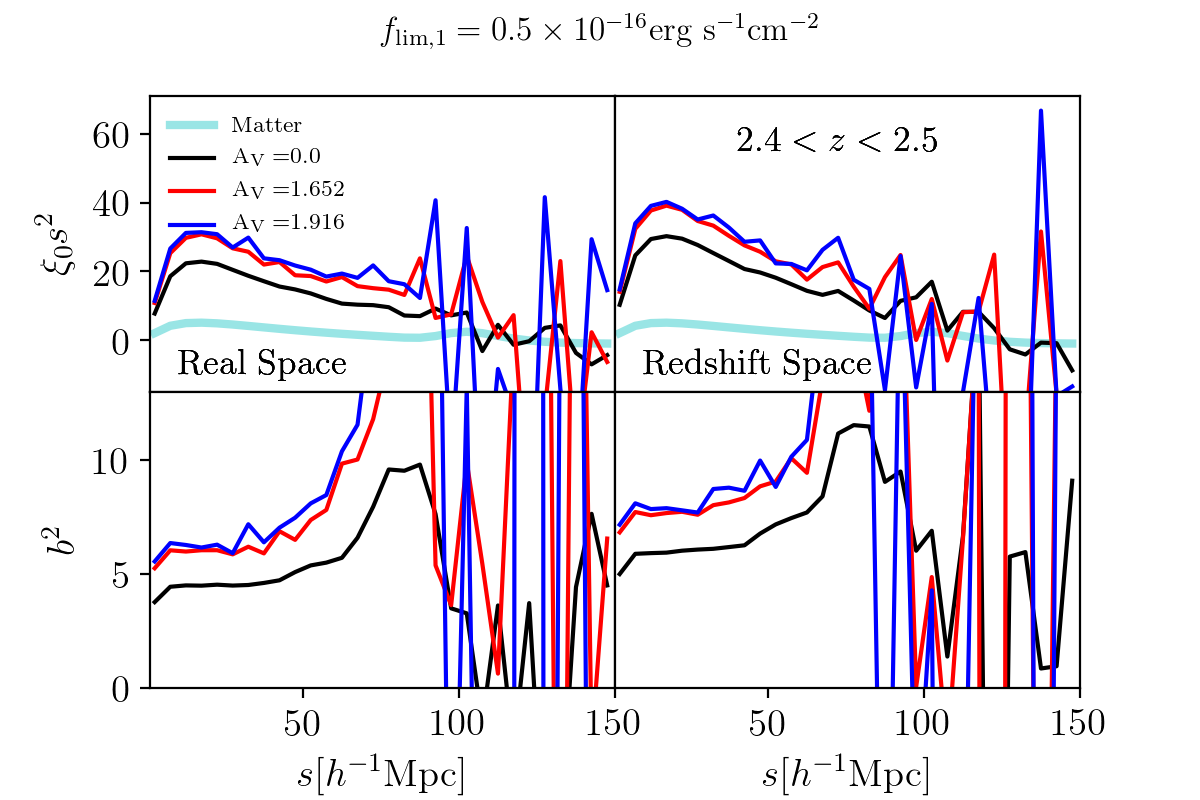}
\caption{The measurement of galaxy correlation function and bias for a few redshift slices of [OIII] galaxies, as labeled in the legend. Since these galaxies only have [OIII] emission bright enough, the sample selection is different from the H$\alpha$ galaxies in terms of a single flux limit $f_{\text{lim, 1}}$. The figures display the effect of dust attenuation for galaxies with a [OIII] flux brighter than $0.5\times10^{-16}\mathrm{erg}/\mathrm{s}/\mathrm{cm}^{2}$. The matter correlation function is shown as the thick cyan curve and the galaxy bias is measured in the bottom panels.}
\label{fig:O3_clustering}
\end{center}
\end{figure*}

\begin{figure*}
\begin{center}
\includegraphics[width=18.5cm]{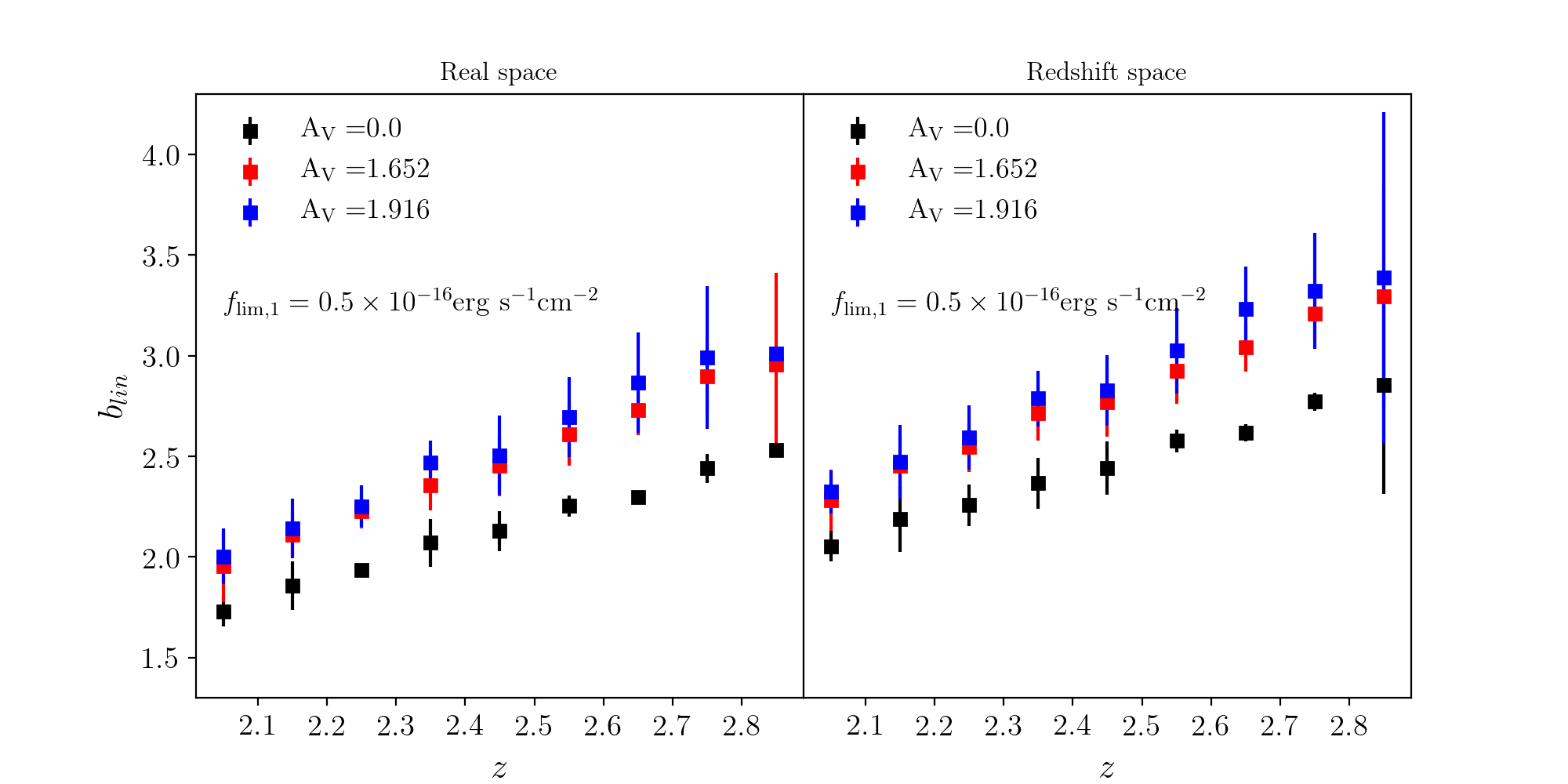}
\caption{Linear bias of the [OIII] galaxies estimated as a function of redshift for both real and redshift space. The galaxies are chosen with $f_{\text{lim, 1}}=0.5\times10^{-16}\mathrm{erg}/\mathrm{s}/\mathrm{cm}^{2}$ and the colors are coded for dust attenuation. The result indicates a significant impact of the dust attenuation, compared with the low-redshift H$\alpha$ galaxies. The measurement of galaxy bias is close to a linear relation with respect to redshift.}
\label{fig:O3_bias}
\end{center}
\end{figure*}

\begin{figure*}
\begin{center}
\includegraphics[width=18.5cm]{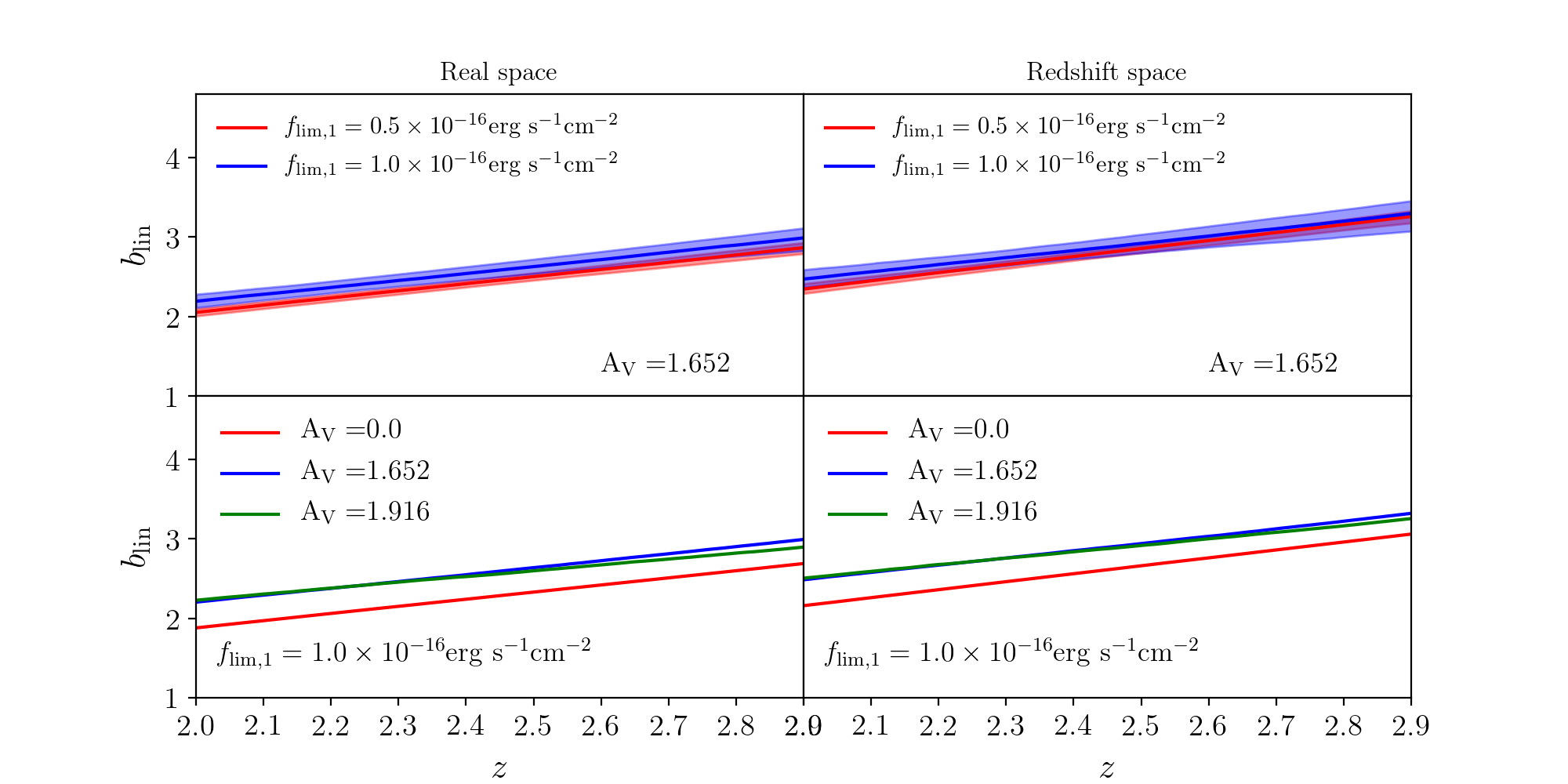}
\caption{Linear bias of [OIII] galaxies as a linear function of redshift using clustering measurement within scales of $10<s<50h^{-1}$ Mpc. Both real (left) and redshift (right) space results are shown. The figure also displays the effect of the flux limit of the emission line and dust attenuation. \textbf{Top:} Galaxies are selected with the WISP-based dust model and colors stand for the flux limit. \textbf{Bottom:} effect of the dust model for galaxies with [OIII] emission line flux brighter than $1.0\times10^{-16}\mathrm{erg}/\mathrm{s}/\mathrm{cm}^{2}$, i.e. the 6.5$\sigma$ nominal depth of Roman. The flux limit has a direct impact on the galaxy bias with brighter galaxies are more biased. The two dust models give consistent estimate of the galaxy bias, similar to the H$\alpha$ galaxies at low redshift. The shaded area represents inner 68\% distribution based on a MCMC test.}
\label{fig:O3_bias_fit}
\end{center}
\end{figure*}

\begin{figure*}
\begin{center}
\includegraphics[width=18.5cm]{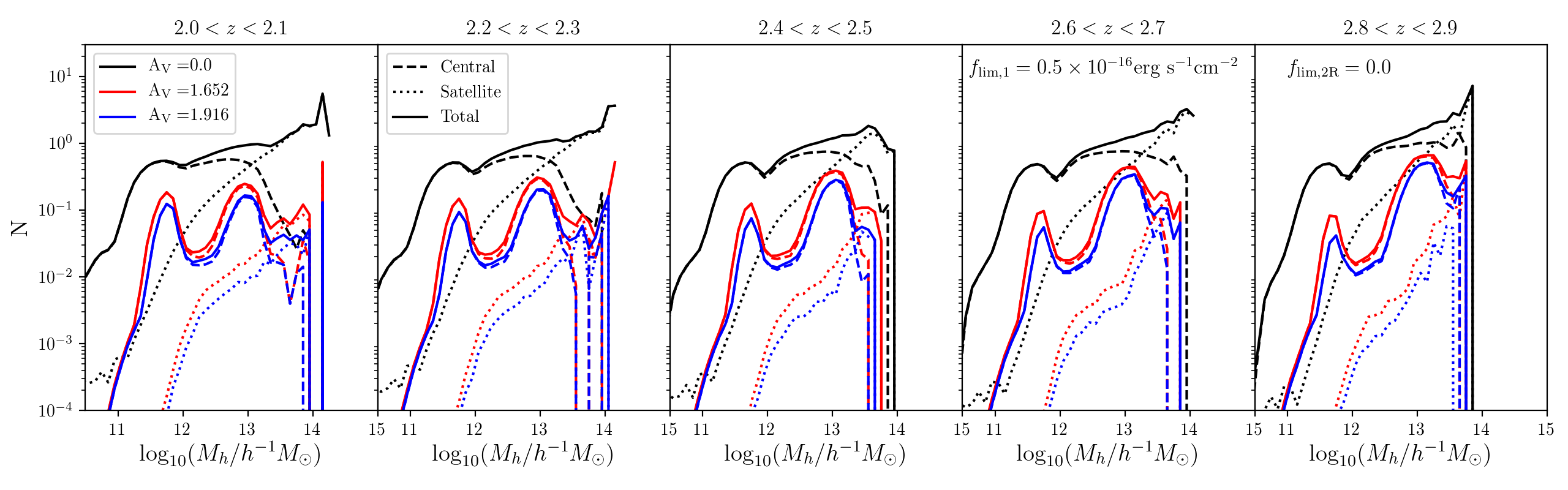}
\includegraphics[width=18.5cm]{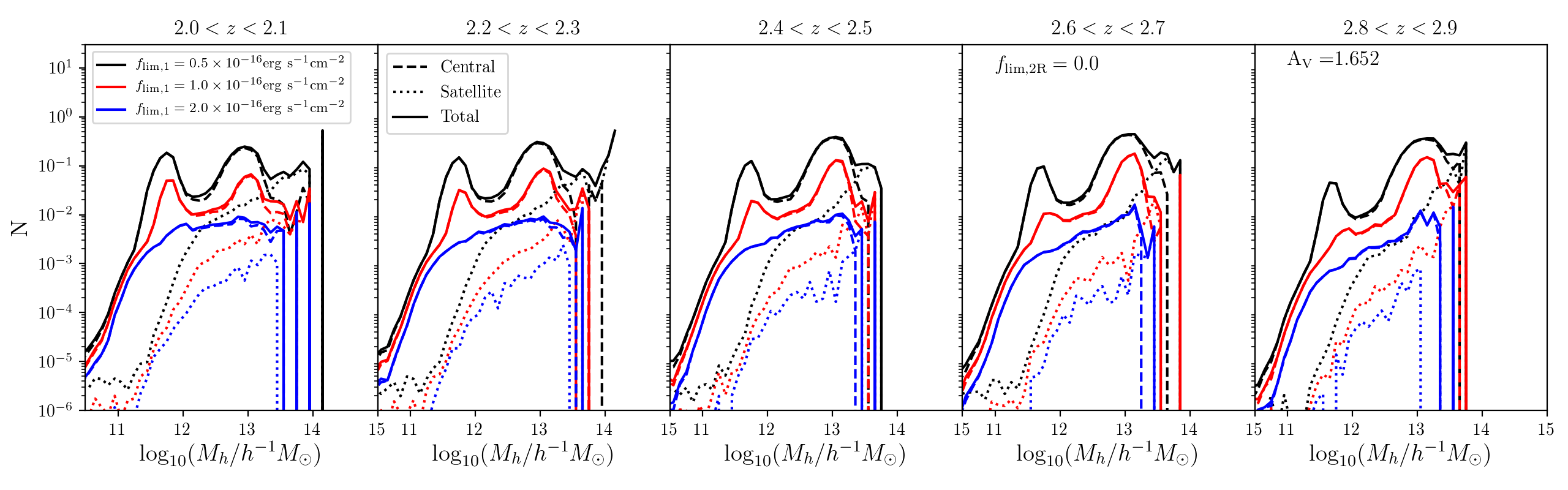}
\caption{Halo Occupation Distribution (HOD) of the [OIII] galaxies for redshift slices as indicated in each panel. The effect of limiting the line flux and dust model is also shown. \textbf{Top:} The effect of dust model for galaxies with [OIII] flux higher than $0.5\times10^{-16}\mathrm{erg}/\mathrm{s}/\mathrm{cm}^{2}$. \textbf{Bottom:} The effect of emission line flux limit with dust model calibrated based on WISP number counts. We present the occupation of centrals (dashed), satellites (dotted) and the total (solid) respectively.}
\label{fig:HODz2}
\end{center}
\end{figure*}

In this section we present the linear bias measurements for galaxies at $z>2$. Note that for this paper, the sample selection for these high redshift galaxies ($z>2$) is different from the lower redshift sample ($z<2$); only the [OIII] line flux is used, 
as $f_{\text{lim, 1}}$, to define the flux-selected samples. The [OII] line can be used as the second emission line for robust redshift determination for the Roman $z>2$ ELG sample, which will be implemented in future work pending improvement of [OII] line flux predictions from Galacticus.

In the top rows of Figure \ref{fig:O3_clustering}, we present the correlation function of the galaxies with a few example redshift slices. The bias measurement $b^{2}$ is shown in the bottom row. We find that the BAO peak can be recovered with these sparser samples, compared with the $z<2$ galaxies. However, due to the lower number density, the clustering measurement becomes noisy and the BAO signal is erased to certain extent. This can be improved by enlarging the redshift bins in the analysis to include more galaxies in the measurement of the two point correlation function. The ratio of the galaxy and the dark matter correlation functions also shows a close to constant behavior at scales $10<s<50$ $h^{-1}$Mpc, similar to the H$\alpha$ galaxies and thus enables an estimate of a constant bias on linear scales.

In Figure \ref{fig:O3_bias}, we present the bias measurement for [OIII] ELGs using the same method as in previous section. The galaxies are selected if the [OIII] line flux is higher than $0.5\times10^{-16}\mathrm{erg}/\mathrm{s}/\mathrm{cm}^{2}$. The stronger dust attenuation (higher $A_{V}$) selects brighter and more massive galaxies, which increases the galaxy bias as expected. Given the estimated uncertainty, the two dust models give consistent result within $1\sigma$. 
Compared with the H$\alpha$ ELGs, the two dust models increase the bias estimates significantly for the [OIII] ELGs, due to the fact that dust imposes more attenuation on [OIII] emission than H$\alpha$.
The distribution of the measurements in the figure also presents a linear relation of galaxy bias with redshift. We fit with a linear relation as introduced in Section \ref{sec:bias_Halpha} and present the result in Figure \ref{fig:O3_bias_fit}. We note that the flux limit has a stronger effect on the bias of the $z>2$ galaxies than the $z<2$ galaxies; the bias can change by roughly 20\%. The nominal depth for Roman galaxy survey of flux above $1\times10^{-16}\mathrm{erg}/\mathrm{s}/\mathrm{cm}^{2}$ is able to observe a significant number of [OIII] emitting and highly biased galaxies. This can provide robust measurement for the clustering signal to infer cosmological information. Compared with the galaxies at $z<2$, these [OIII] galaxies are more biased due to the early phases of the dark matter evolution and the redshift dependence of dark matter halo bias. The fitting result of the linear bias model is summarized in Table \ref{tab:OIII}.

We measure the HOD of these [OIII] galaxies to better understand their distribution within dark matter halos and present the result in Figure \ref{fig:HODz2} for a few redshift slices. The prominent feature is similar to that of the H$\alpha$ galaxies at $z<2$. The central occupation shows a clear double-peak behavior as a function of halo mass. Either it is caused by physical reasons of mass loss of dark matter halos, or numerical issues in the simulation, we will investigate this in future work. Similar to the $z<2$ galaxies, the satellite occupation is also close to a power-law form, but with some break at high redshift. 

\begin{table}
\centering
\begin{tabular}{llll}
\hline
Real Space &  $a$ & $b$   \\
\hline
$f_{\text{lin, 1}}=0.5, A_{V}=0.0$  & $0.842\pm0.028$   & $0.062\pm0.068$ \\
$f_{\text{lin, 1}}=0.5, A_{V}=1.652$  & $0.928\pm0.093$   & $0.185\pm0.218$ \\
$f_{\text{lin, 1}}=0.5, A_{V}=1.915$  & $0.91\pm0.128$   & $0.261\pm0.291$ \\
\hline
$f_{\text{lin, 1}}=1.0, A_{V}=0.0$  & $0.906\pm0.032$   & $0.063\pm0.078$ \\
$f_{\text{lin, 1}}=1.0, A_{V}=1.652$  & $0.917\pm0.183$   & $0.37\pm0.407$ \\
$f_{\text{lin, 1}}=1.0, A_{V}=1.915$  & $0.75\pm0.27$   & $0.716\pm0.612$ \\
\hline
Redshift Space &  $a$ & $b$   \\
\hline
$f_{\text{lin, 1}}=0.5, A_{V}=0.0$  & $0.923\pm0.067$   & $0.196\pm0.169$ \\
$f_{\text{lin, 1}}=0.5, A_{V}=1.652$  & $1.03\pm0.113$   & $0.276\pm0.27$ \\
$f_{\text{lin, 1}}=0.5, A_{V}=1.915$  & $1.07\pm0.124$   & $0.226\pm0.278$ \\
\hline
$f_{\text{lin, 1}}=1.0, A_{V}=0.0$  & $1.018\pm0.047$   & $0.116\pm0.12$ \\
$f_{\text{lin, 1}}=1.0, A_{V}=1.652$  & $0.98\pm0.251$   & $0.49\pm0.558$ \\
$f_{\text{lin, 1}}=1.0, A_{V}=1.915$  & $0.771\pm0.318$   & $0.99\pm0.725$ \\
\end{tabular}
\caption{The flux limit $f_{\text{lin}, 1}$ is in unit of $\times10^{-16}\mathrm{erg}/\mathrm{s}/\mathrm{cm}^{2}$. The fitting result for the galaxy bias as a linear function of redshift: $b_{\text{lin}}(z)=az+b$, estimated from clustering measurement in both real and redshift space for [OIII] galaxies within $2<z<3$.}
\label{tab:OIII}
\end{table}

\subsection{Comparison of HOD with eBOSS}

\begin{figure*}
\begin{center}
\includegraphics[width=17.5cm]{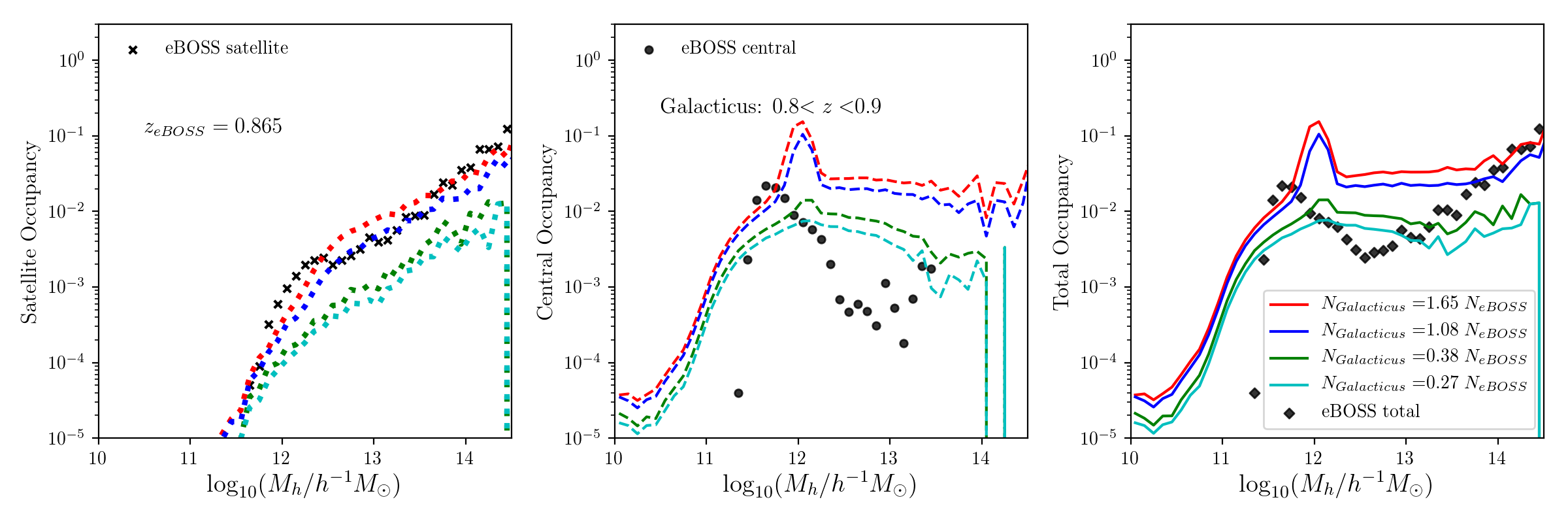}
\caption{Comparison of HOD measurements with eBOSS. \textbf{Left:} occupancy for satellites; \textbf{Middle:} occupancy for centrals, \textbf{Right:} occupancy for total galaxies. Using Roman simulation, we apply flux cut $f_{\mathrm{lim}}=5\times10^{-16} \mathrm{erg s}^{-1}\mathrm{cm}^{-2}$ and $10\times10^{-16} \mathrm{erg s}^{-1}\mathrm{cm}^{-2}$, and dust models $A_{v}=1.6523$ and $A_{V}=1.9156$ to define four galaxy mocks within $0.8<z<0.9$. The number densities of these mocks are 1.65, 1.08, 0.38 and 0.27 times that of eBOSS ELG respectively and their results are marked by different colors. The eBOSS measurements are from \citet{Avila_2020}.}
\label{fig:HOD_eBOSS}
\end{center}
\end{figure*}

In order to further investigate whether our Galacticus simulation can make reasonable predictions for HODs, we compare the HOD results with the latest eBOSS ELG measurement (\citealt{Avila_2020}). The eBOSS ELG program creates a catalog of thousands of galaxies within the redshift range of $0.6<z<1.1$, selected using the DECaLS photometric survey. The finalized sample has an average redshift $z=0.865$ with a number density of $n_{\mathrm{eBOSS}}=2.187\times10^{-4}(\mathrm{Mpc}/h)^{-3}$. At this redshift, the Roman HLSS can only observe H$\alpha$ emission due to the wavelength range of its grism (see Figure \ref{fig:redshift_range}). Therefore we use the H$\alpha$ flux to define our galaxy mock. We apply flux limit $f_{\mathrm{lim}}=5\times10^{-16}\mathrm{erg s}^{-1}\mathrm{cm}^{-2}$ and $10\times10^{-16}\mathrm{erg s}^{-1}\mathrm{cm}^{-2}$ with two dust models $A_{v}=1.6523$ and $A_{V}=1.9156$. This give us four different galaxy samples with number densities 1.65, 1.08, 0.38 and 0.27 times that of eBOSS ELG. We measure their HODs and compare with the eBOSS measurement in Figure \ref{fig:HOD_eBOSS}. Although our galaxy sample has different target selection than eBOSS which uses [OII] doublet to identify galaxy redshift, our prediction of HOD has similar amplitude when the number density is close to that of eBOSS. The satellite occupancy shows excellent agreement in terms of the shape and amplitude. The halo mass dependence can be described by a power-law at high mass end. The central occupancy in both Galacticus and eBOSS shows a similar shape with a peak at intermediate mass scale, with eBOSS peaking at slightly lower mass scale. In addition, the overall amplitude of the central occupancy of Galacticus is higher and its shape flattens at high mass end instead of dropping quickly. This discrepancy can be caused by a combination of factors: the difference in the selection algorithms of Roman GRS and eBOSS, the calibration of Galacticus for the parameters governing star formation history and galaxy formation, the dust models used in the analysis and so on.

\subsection{A practical fit of ELG bias at 1 < z < 3}

\begin{figure*}
\begin{center}
\includegraphics[width=17.5cm]{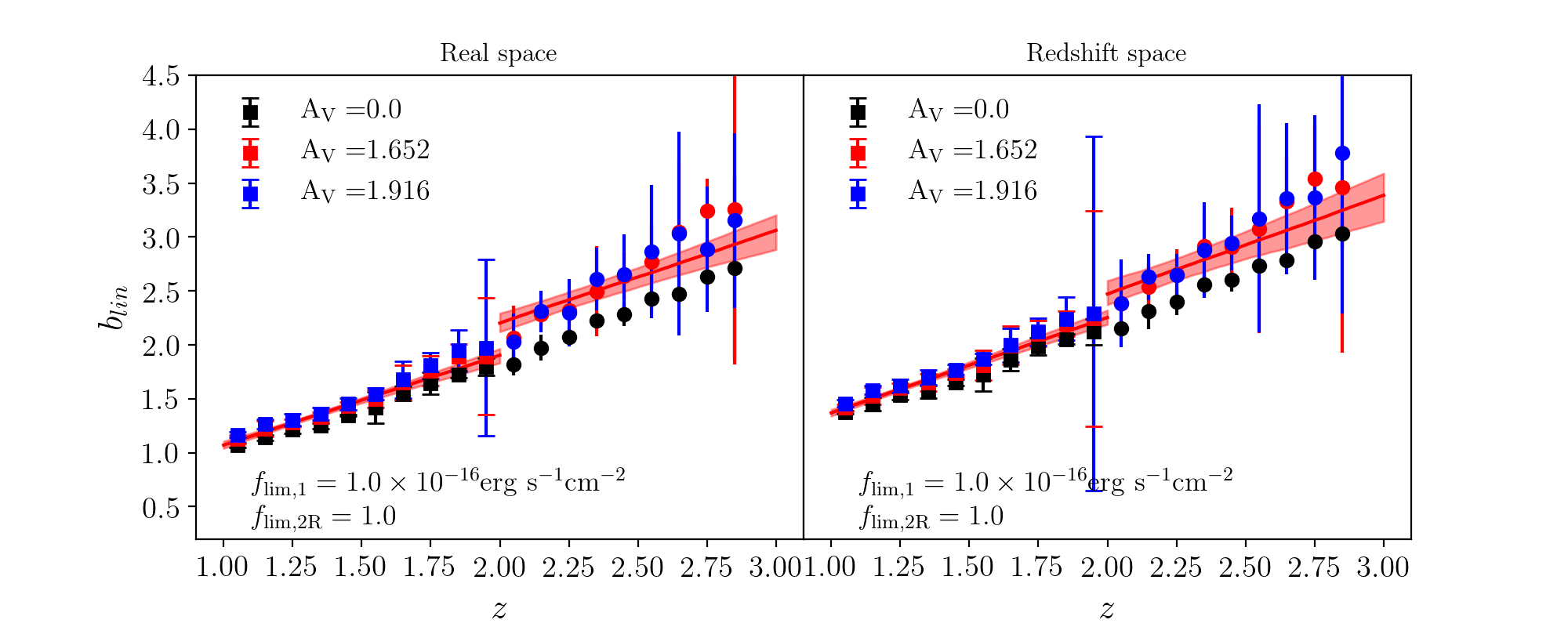}
\caption{The linear bias measurement for the entire redshift range of Roman HLSS. The H$\alpha$ galaxies are selected by two emission lines with flux limits $f_{\text{lim,1}}$ and $f_{\text{lim,2R}}$, while [OIII] galaxies are selected by $f_{\text{lim,1}}$ only. The result shows measurements with dust-free ($A_{\text{V}}=0$) and two dust models. The red lines are a linear fit for measurements with $A_{\text{V}}=1.652$ and the shaded area is $1\sigma$ uncertainty.}
\label{fig:bias_allz}
\end{center}
\end{figure*}

In Figure \ref{fig:bias_allz}, we present the bias measurement of emission line galaxies for the entire redshift range of Roman HLSS. We apply $f_{\mathrm{lim}}=1\times10^{-16} \mathrm{erg s}^{-1}\mathrm{cm}^{-2}$ and $f_{\text{lin, 2R}}=1.0$ to choose H$\alpha$ galaxies, and $f_{\mathrm{lim}}=1\times10^{-16} \mathrm{erg s}^{-1}\mathrm{cm}^{-2}$ for [OIII] galaxies. Using the measurement with dust model $A_{\text{V}}=1.652$, we perform a linear fit of the bias measurement for H$\alpha$ and [OIII] galaxies respectively, shown as the red line with shaded area. The fitting result can also be found from Table \ref{tab:Halpha_real}, \ref{tab:Halpha_redshift} and \ref{tab:OIII}. In redshift space, we summarize the results as
\begin{eqnarray}
 \text{H$\alpha$ (1<z<2):} & b=0.88z+0.49 \nonumber\\
 \text{[OIII](2<z<3):} & b=0.98z+0.49
 \label{eq:bias}
\end{eqnarray}
Our previous tests show that the practical choices for the dust model and flux limits for the emission lines won't have significant impact on the estimate of the linear bias. Therefore the result quoted above is a reasonable description for future analysis, especially for the investigation of the science forecast of Roman HLSS.

\section{Discussion and Conclusion}

Emission line galaxies are the main targets of many current and future cosmological surveys. The bright nebula emissions due to star formation activity makes the sample selection different from that of the red and massive galaxies at low redshifts. The results based on current observational data are not sufficient to allow simple extrapolation to higher redshifts, thus requires detailed investigations of their spatial distribution based on accurate numerical simulations. In this paper, we study the linear bias of these ELGs from Roman galaxy redshift survey based on clustering measurement and present their redshift evolution for various sample selections and dust models. In particular, we use the Galacticus SAM to perform a large scale galaxy simulation. The model processes all the dark matter merger trees distributed within the 1$h^{-1}$Gpc box of the N-body simulation UNIT (\citealt{Chuang_2019}). We then construct a lightcone catalog using the method in \cite{Kitzbichler_2007}. The parameters of the model are calibrated to match the current observations at high redshifts to ensure that the galaxy simulation is realistic. We used this model to predict the number densities of H$\alpha$ and [OIII] emitters for the Roman galaxy survey in \cite{Zhai_2019MNRAS}. The same model has been used here to produce a 2000 deg$^2$ galaxy mock, consistent with the baseline design of Roman galaxy redshift survey. A galaxy clustering analysis based on this mock catalog is preformed in \cite{Zhai_2020} to forecast the uncertainties of the BAO and RSD measurements. The wavelength range of the Roman grism has a direct impact on the redshift range of each nebula emission line, and constrains the selection of galaxy samples (see Figure \ref{fig:redshift_range}). We have investigated how the clustering of the Roman galaxies depend on the chosen line flux limits.

Depending on the selection criteria of emission line galaxies, we can measure the linear bias of galaxies as a function of redshift with the simulated galaxy catalog and the dust model. We first measure the two-point correlation function of galaxies in real and redshift space and find that the BAO peak on large scales can be recovered for both H$\alpha$ and [OIII] galaxies within the Roman redshift range, although the [OIII] galaxy samples are more affected by shot noise due to the low number density. Taking the ratio of correlation function between galaxies and matter enables the measurement of galaxy bias. The result at scales $10<s<50~h^{-1}$Mpc reveals a roughly constant bias estimate. Thus we use this scale-independent value as the linear bias of galaxies. Deviation of the bias measurement from a constant value at larger scales is noticeable, which is caused by a combination of factors including the non-linear evolution of the BAO signal, the redshift space distortion effect and sample variance due to limited cosmic volume. 

We find that the scale-independent galaxy biases for both H$\alpha$ and [OIII] ELGs are close to a linear function, $b(z)=az+b$, see Eq.(\ref{eq:bias}), consistent with previous results on H$\alpha$ ELGs (\citealt{Merson_2019}), see Table \ref{tab:Halpha_real}, \ref{tab:Halpha_redshift} for H$\alpha$ galaxies, and Table \ref{tab:OIII} for [OIII] galaxies. 
For H$\alpha$ galaxies we have investigated the impact of the line flux limit and dust model on the linear bias measurement, as shown in Figure \ref{fig:Halpha_bias_fit}. We find that the linear bias of ELGs at $1<z<3$ is insensitive to line flux cut or dust attenuation model, consistent with earlier work at $z<2$ \citep{Merson_2019}.
We find that galaxy bias increases with redshift for ELGs, as expected, since higher redshifts correspond to earlier (more biased) phases of galaxy distribution.
As dark mater halos grow with decreasing redshift, they become more populated with galaxies, which reduces the bias factor with which galaxy distribution traces the matter distribution.

In order to better understand the distribution of ELGs within their host dark matter halos, we have performed HOD measurements for the galaxy samples, as well as the halo mass function of the selected galaxies (see Appendix \ref{appsec:hmf}). The noticeable feature is the double peak for the central occupancy. The second peak at high mass end is likely caused by the mass loss of halos during evolution. However the current model is not able to identify whether this mass loss is physical or due to numerical artifacts, therefore we leave it for future work. On the other hand, the satellite occupancies for both H$\alpha$ and [OIII] galaxies are close to a power-law form, with the tendency of a break at the high mass end. This can enable a simple parameterization for practical application in the analysis of large scale structure.

The Roman galaxy redshift survey will suffer from the usual systematic effects of slitless spectroscopy \citep{Faisst_2018, Martens_2019}, such as line misidentification and spectral overlap, although to a lesser degree compared to Euclid, thanks to the higher spectral resolution and wider wavelength range of the Roman grism compared to the Euclid red grism (the Euclid blue grism will not be used in the wide survey). In future work, we will study the survey completeness and purity for the Roman galaxy redshift survey, and their impact on the observed galaxy sample and the galaxy clustering analysis. 

The HOD measurement of the Roman galaxy sample builds a straightforward connection between galaxies and dark matter halos, in terms of the halo mass being the only parameter. However the secondary halo properties other than halo mass can also impact the clustering signals of galaxies. This assembly bias phenomena has been reported in researches based on numerical simulation, see e.g. \cite{Gao_2005, Wechsler2006}. Some of the latest studies for ELGs find that the secondary properties of dark matter halos can affect the distribution of ELGs and thus the cosmological measurement based on large scale structure like BAO peak (\citealt{Jimenez_2020}). The SAM employed in our simulation can output detailed properties of galaxies, and their host halos. This can build a more accurate connection between galaxies and dark matter halos by using information of both internal and external halo properties, and help minimize the systematics in the cosmological inference.

We have presented linear bias and HOD measurements for ELGs at $1<z<3$ in this paper, which are key inputs to the realistic forecast of dark energy and cosmological constraints from possible Roman galaxy redshift surveys. These in turn can be used to optimize the observing strategy for Roman. Our results are similarly useful to other ongoing or future galaxy surveys that use ELGs to trace cosmic large scale structure.


\section*{Data Availability}
The original dark matter halo catalogs are available from the UNIT simulation website. The galaxy mocks are available by request. A public webpage presenting the mocks will be available at a later time.

\section*{Acknowledgements}

ZZ thanks Santiago Avila for providing their HOD measurement of the eBOSS ELG sample. This work is supported in part by NASA grant 15-WFIRST15-0008, Cosmology with the High Latitude Survey Roman Science Investigation Team (SIT). GY would like to thank MICIU/FEDER (Spain)  for financial support under project grant  PGC2018-094975-B-C21.
The UNIT simulations have been done in the MareNostrum Supercomputer at the Barcelona Supercomputing Center (Spain) thanks to the  cpu time awarded by PRACE under project grant number 2016163937. This work used the Extreme Science and Engineering Discovery Environment (XSEDE), which is supported by National Science Foundation grant number ACI-1548562 (\citealt{XSEDE_2014})

\rm{Software:} Python,
Matplotlib \citep{matplotlib},
NumPy \citep{numpy},
SciPy \citep{scipy}, George \citep{george_2014}, emcee \citep{Foreman-Mackey_2013}, Hankel \citep{Murray2019}.


\bibliographystyle{mnras}
\bibliography{emu_gc_bib,software}

\appendix

\section{ELG halo mass function}\label{appsec:hmf}

\begin{figure}
\begin{center}
\includegraphics[width=8.5cm]{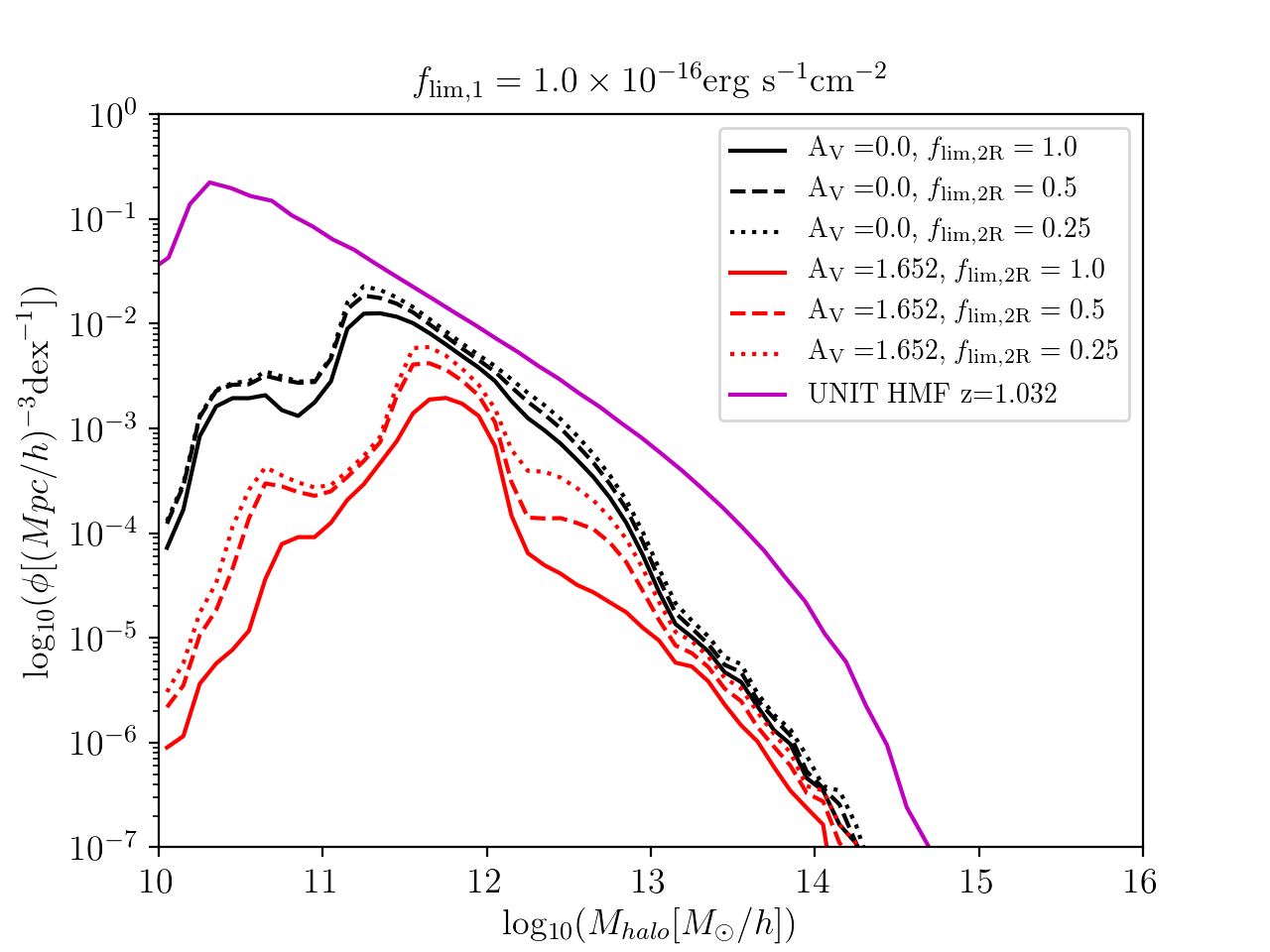}
\caption{Illustration of the halo mass function of the Roman ELGs within $1.0<z<1.1$. Galaxies are selected with the criteria including flux limits and dust model as shown in the legend. The solid purple line is the halo mass function from the original UNIT simulation at $z=1.032$. }
\label{fig:hmf}
\end{center}
\end{figure}

In order to better understand the distribution of galaxies within dark matter halos, we compute the halo mass function (HMS) of the Roman ELGs, i.e. the number density of dark matter halos that host Roman galaxies. Figure \ref{fig:hmf} shows an example for galaxies within $1.0<z<1.1$, with different selection algorithms including the flux limits on the emission lines and dust models. The lower amplitude for the Roman galaxies is mainly dominated by the selection algorithms. The flux limit can remove faint galaxies which are likely to live in less massive halos. The requirement of both H$\alpha$ and [OIII] emission lines and the galaxy formation physics impact the decrease of the HMS at high mass end. 

\bsp	
\label{lastpage}
\end{document}